\documentclass[twocolumn,tighten,times]{aastex61}

\usepackage{color}
\usepackage{enumerate}
\usepackage{esvect} 
\usepackage[section]{placeins} 
\usepackage{ulem} 
\usepackage{array}
\usepackage{multirow}
\usepackage{verbatim}
\usepackage{graphicx}
\usepackage{amsmath,bm}
\usepackage{txfonts}

\hypersetup{linkcolor=red,citecolor=blue,filecolor=cyan,urlcolor=magenta}

\newcommand\aastex{AAS\TeX}

\def\deg{{}^{\circ}}

\shorttitle{\aastex\ On the close encounters between Plutinos and Neptune Trojans}
\shortauthors{Dong \& Zhou}

\begin{document}

\title{On the close encounters between Plutinos and Neptune Trojans: \\\uppercase\expandafter{\romannumeral1}.  Statistic analysis and theoretical estimations}

\author{DONG Cheng-Yu}
\affiliation{School of Astronomy and Space Science, Nanjing University, Nanjing 210046, China}

\author{ZHOU Li-Yong}
\altaffiliation{zhouly@nju.edu.cn}
\affiliation{School of Astronomy and Space Science, Nanjing University, Nanjing 210046, China}
\affiliation{Key Laboratory of Modern Astronomy and Astrophysics in Ministry of Education, Nanjing University, Nanjing 210046, China}
\correspondingauthor{ZHOU Li-Yong}



\begin{abstract}

Close encounters (CEs) between celestial objects may exert significant influence on their orbits. The influence will be even enhanced when two groups of celestial objects are confined in stable orbital configurations, e.g. in adjacent mean motion resonances (MMRs). Plutinos and Neptune Trojans, trapped in the 2:3 and 1:1 MMRs with Neptune respectively, are such examples. 
As the first part of our investigation, this paper provides a detailed description of CEs between Plutinos and Trojans and their potential influences on the Trojans' orbits. Statistical analyses of CE data from numerical simulations reveal the randomness lying in the CEs between the two planetesimals. The closest positions of CEs distribute symmetrically inside the given CE region and no particular bias is found between the positive and negative effects on the orbital elements of Trojans. Based on the Gaussian approximation on the distribution of the velocity orientation of Plutino, and the integral derivatives of Gaussian perturbation equations, a theoretical method is built to estimate the CE effects. To further verify the randomness of CEs, a Monte Carlo approach is applied, and it generates distribution features consistent with the numerical results. In summary, CEs brought by realistic Plutinos exert impartial effects and tiny total influence on the orbital elements of Trojans. However, driven by the random walk mechanism, tiny effects may accumulate to a prominent variation given sufficient CEs, which will be discussed in the accompanying paper.

\end{abstract}


\section{Introduction}

Neptune Trojans refer to a cluster of minor objects that orbit the Sun and locate around the stable Lagrangian points of Neptune, dynamically trapped in the 1:1 mean motion resonance (MMR) with Neptune. Among the so far discovered Neptune Trojans\footnote{IAU: Minor Planet Center, \url{http://www.minorplanetcenter.net/iau/lists/NeptuneTrojans.html}}, more than a half posses high inclinations. 
The dynamical stability and the driving mechanism of these highly inclined orbits have long become perplexing problems.

Secular perturbations from giant planets may significantly shape the dynamic structure of the resonant region, and indeed several works managed to locate stable regions of Neptune Trojans around high inclinations \citep{2003MNRAS.345.1091M, 2007MNRAS.382.1324D, 2008CeMDA.102...97D, 2009MNRAS.398.1217Z, 2011MNRAS.410.1849Z}, but the specific mechanism which effectively elevates the inclination of Trojans remains ambiguous. The Nice Model \citep{2005Natur.435..462M} proposes that the chaotic capture in Jovian Trojan region during the primordial migrations of giant planets will remarkably increase the inclination of Trojans. Similar mechanism is applicable to Neptune Trojans as well \citep{2004MNRAS.347..833B, 2007A&A...464..775L}. However, as proposed by \citet{2009AJ....137.5003N}, the current capture efficiency is insufficient to explain the 4:1 ratio between Neptune Trojans with high inclinations and low inclinations \citep{2006DPS....38.4403S}. Besides, it is rather unreliable that the early formed distribution structure can survive the lengthy evolution later.  
   Certainly, the artificiality of the model is an insecure factor either.
     
Given the above considerations, we draw our attention to the interactions prompted by adjacent asteroid families like Plutinos. Plutinos are classified as a prime subpopulation of Trans-Neptunian objects (TNOs), and trapped in a 3:2 MMR with Neptune. With the semi-major axes locating around 39 AU and eccentricities extending up to 0.3, Plutinos can cross the orbit of Neptune and meet Trojans from time to time. \citet{2009A&A...508.1021A} suggest that the significant orbital overlap will lead to frequent close encounters between Plutinos and Trojans. This event, bound by stable resonant configurations, is different from the interior close encounters among either population, thus may be an extra force to effectively shape the physical and orbital characteristics of the two interacting clusters. The idea then comes up that the frequent communications with Plutinos may play a significant role in the formation of high inclinations of Trojans. If standing up, this mechanism will be an efficient way to continually and inherently elevate the inclinations of Trojans, thus contributes to the present distributions.
   
       \begin{figure}[htbp]
    \centering
    \includegraphics[width=1.0\hsize]{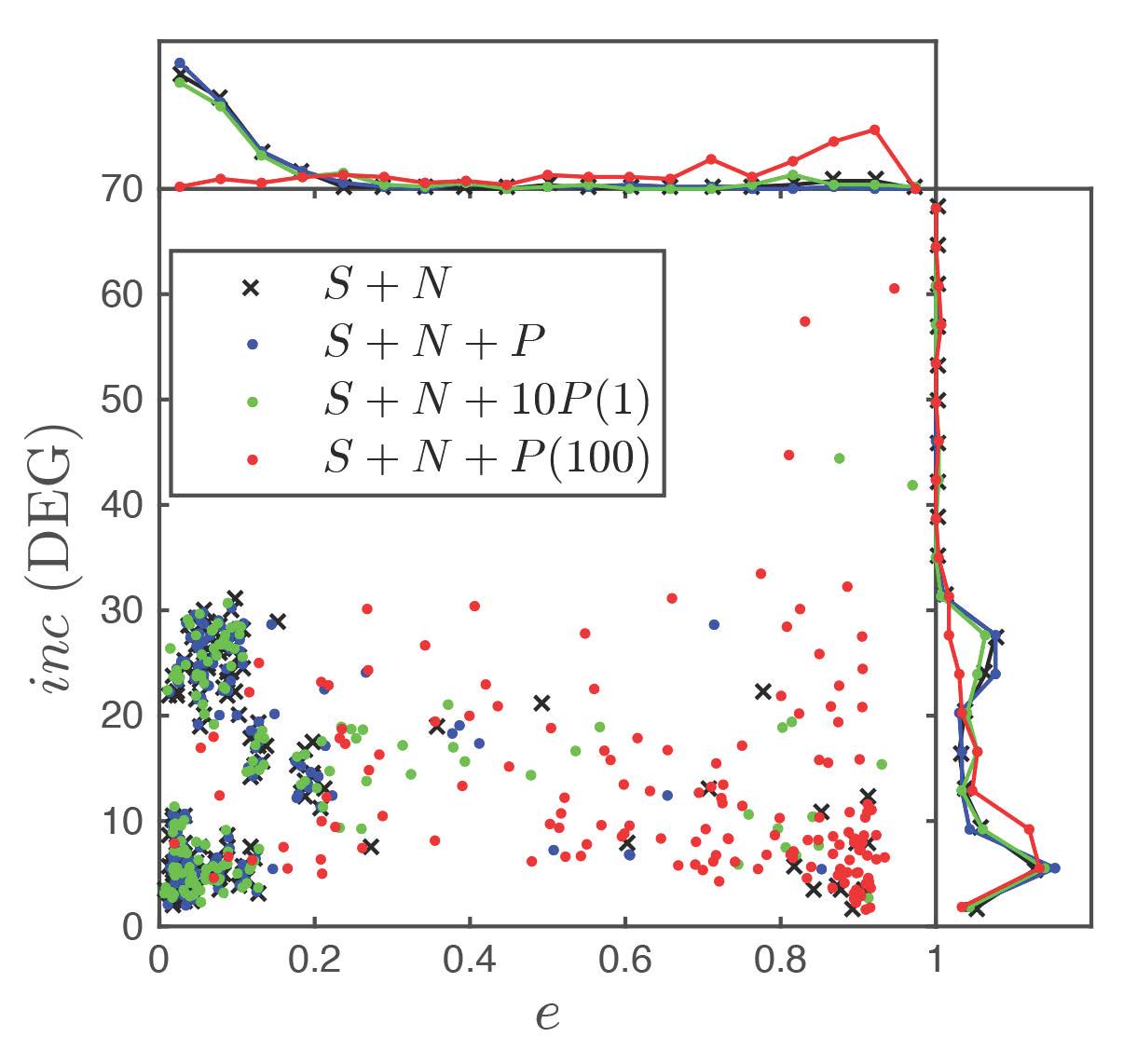}
      \caption{The orbital element distribution of a fictitious population of Trojans after 1Gyr numerical integration. Different colors denote results from different models, namely the model consisting of only the Sun and Neptune ($S+N$), the Sun, Neptune and Pluto ($S+N+P$), the Sun, Neptune and 10 fictitious Plutinos with Pluto mass each ($S+N+10P(1)$), the Sun, Neptune and Pluto with fictitious 100 Pluto mass ($S+N+P(100)$). 
      Trojans with large eccentricities are scattered away from the resonant region, with their semi-axes reaching up to 400 AU.
      The attached plots show the histograms of the corresponding orbital elements.}
         \label{intro1}
    \end{figure} 
    
   Simply and directly, we test this idea by simulating the motion of a fictitious population of massless Trojans, using the state-of-art numerical integrator SyMBA \citep{2000AJ....120.2117L}. The Trojans are cloned from the known ones. As the minimum requirements to sustain Trojans, the standard model consists of the Sun and Neptune, while other models also include Pluto or several Plutinos. All models are integrated on the same population of Trojans simultaneously in order to reveal the differences. In Fig.\,\ref{intro1} , we show the orbital element distribution of Trojans perturbed by Pluto after 1Gyr evolution, which barely shows any difference from the standard case. The histograms of inclination almost coincide with each other. A further model including 10 fictitious Plutinos each with Pluto mass still brings minute effect. Nevertheless, when we introduce a fictitious Pluto with 100 times the realistic Pluto mass (about 0.2 Earth mass), the Trojans are scattered away from the Neptune orbit, indicating that Plutinos are indeed able to reach and influence the orbits of Trojans. 
   Surely that under realistic situations, the total mass of Plutino population could never be as high as 100 Pluto mass. But it causes our major concern about the number of Plutinos. Currently there are about one hundred identified Plutinos. Could the cumulative effect of such number of Plutinos significantly perturb the orbits of Trojans through incessant close encounters, though their total mass is relatively low? Via pure numerical simulation, we then need to include all these known bodies simultaneously, which requires huge computing resources. And these bodies may still be just a small portion of the population in reality. It is hard to determine the exact number of bodies to be included in the simulation that is convincing enough to reflect the realistic effect. Besides, a pure numerical method can not be generalized for sure. When facing a new but similar problem associated with the communication between overlapping resonances, we still need to cover all of the bodies and perform numerical simulations, which is redundant effort but hardly touches the physical essence.

Apart from the direct numerical integrations, statistical analyses and theoretical estimations are always alternative ways. With regard to the close encounters in the solar system, numerous works focus on the circumstances that minor objects like asteroids and comets encounter giant planets, and resolve to unravel the consequent effect on the orbital elements or energy \citep{1969AJ.....74..735E, 1988Icar...75....1G, 1990CeMDA..49..111C, 2013A&A...550A..85C}. Statistical analyses and semi-analytical methods are frequently applied in these explorations. 
   
More violent processes refer to collisions, and as a prerequisite for general analysis of collisional evolution, an estimate of the collision probability has been studied in various ways. Classical and practical theories include but not limited to \citet{1967JGR....72.2429W}, \citet{1982AJ.....87..184G}, \citet{1993GeoRL..20..879B} and \citet{1998Icar..136..328D}. Pertinent methods were applied to the collisions among groups of minor objects including planetary embryos \citep{1997Icar..128..429W}, Main Belt asteroids \citep{1992Icar...97..111F}, Hildas \citep{2001A&A...366.1053D}, Jupiter Trojans \citep{1996Icar..119..192M} and near-Earth objects\citep{2004Icar..170..295S}. With the continuous discovery of TNOs, more interest was devoted to the collisions among relevant subpopulations \citep{1995AJ....110..856S, 1997Icar..125...50D, 2003Icar..162...27T, 2005Icar..174..105K, 2013A&A...558A..95D}. 
Numerical methods including Monte Carlo simulations are frequently applied in these explorations.
   
Following the same route, in this paper we will deal with the close encounters between Neptune Trojans and Plutinos, which happen much more frequently than collisions, but have been rarely studied beforehand. Since the pure numerical model is incapable or not convincing enough to reveal the effect of close encounters, we shall start from the simplest case, including only one Plutino and one Trojan. By controlling key variables, this idea allows us to explore the important relations between the geometries or effects of close encounters, and the physical or orbital characteristics of either interacting planetesimals. Thanks to the stable orbital configuration offered by overlapping resonances, concise and straightforward theoretical estimations are available to explain these relations, which gives us a better understanding of communications in such case.
Furthermore, by introducing the random walk model to combine the theoretical distributions of the close encounter effect and the well-developed theoretical formula on the encounter frequency, we can actually estimate the cumulative effect contributed by single planetesimal or a group of planetesimals with different characteristics. The effect of realistic population of Plutinos exerted on Trojans, or at least the upper limit,  can then be finally approached with enough confidence. And obviously this analytical route can be generalized to an arbitrary problem associated with overlapping resonances, such as the communication between the Hilda group and Jupiter Trojans in the solar system, and similar cases in the extrasolar systems. The latter part is discussed in the accompanying paper.
   
Therefore, the outline of this paper goes as follows. In Sect. \ref{Stat}, numerical simulations will be implemented for cases including only one Trojan and one Plutino. Statistical analyses on the frequency and geometries of the close encounters, along with available theoretical explanations and estimations will be done at the same time. Those who are not interested in the statistical tools or the derivation of the analytical formula can choose to skip these details. In Sect. \ref{RanSim}, Monte Carlo simulations are implemented to further justify the features of the close encounters. Results and implications are summarized in Sect. \ref{conclu}.

\section{Statistic behaviors of observed planetesimals}\label{Stat}

In this section, we statistically explore the close encounters (CEs) between Plutinos and Neptune Trojans. The first question comes that whether or not there lies a bias in the intermittent influence on Trojan by Plutinos, which will probably result in the variation of the orbital elements of Trojan. To address that, we first need to detect CEs in the numerical simulations. The number of detected CEs can indicate the intensity of such events for different orbital characteristics. Meanwhile, by measuring the relative distance and directions in CEs between Plutino and Trojan, we can infer that if there is any bias lying in the interaction. More directly, we can collect the inclination and eccentricity change of Trojan caused by each CE, whereby the influence bias will be revealed clearly. In addition, the bias may concentrate within certain timespan, whereupon the time that CE occurs matters as well. Throughout these analyses, we are able to explore the geometries and dynamic effect of the CEs, as well as their crucial relations with each other, and more importantly, with the orbital elements of Plutinos and Trojans. Such relations will allow us to build a rigorous analytical formula to derive the magnitude and distribution of CE effect from the orbital characteristics of interacting planetesimals.

We pick several realistic planetesimals into our simulations as examples. The orbital elements are listed in Table~\ref{orbel} for reference (from \citealt{2009A&A...508.1021A}). These examples are chosen in order to represent different inclination levels, based on the supposition that the inclination affects CE effect the most among all six orbital elements. The eccentricity of Plutino example should be sufficiently large to allow its orbit to approach Trojan, that is, about $0.25$, which is a common value among realistic Plutinos. 
    
Note that, throughout this paper, the discussion will be focused on these examples, including the theoretical derivations. Nevertheless, our theoretical method will not be limited to these particular cases. It will be demonstrated in the accompanying paper \citep{2017_2} that this theoretical method is actually applicable to other conditions, despite the specific initial physical or orbital characteristics of the interacting planetesimals. Based on this, we may further estimate the CE effect of arbitrary interacting planetesimals or even a group of them, thus approaching our expectation.
    
   \begin{table}[htbp]
      \caption{Orbital elements of several Plutinos (2004 UP10 and 2006 RJ103) and Neptune Trojans (1999 CE119 and 2001 FU172) involved (JD 2454200.5).}
         \label{orbel}
     $$          \begin{array}{p{0.18\linewidth}p{0.10\linewidth}p{0.1\linewidth}p{0.1\linewidth}p{0.12\linewidth}p{0.12\linewidth}p{0.12\linewidth}l}
            \hline\hline
             & $a$ & $e$ & $i$ & $M$ & $\omega$ & $\Omega$ \\ 
             Objects & $\rm(au)$ &  & $(\mathrm{\deg})$ & $(\mathrm{\deg})$ & $(\mathrm{\deg})$ & $(\mathrm{\deg})$ \\
            \hline
            \noalign{\smallskip}
            2004\,UP10 & 30.099 & 0.025 & 1.4 & 334.1 & 2.2 & 34.8 \\
            2006\,RJ103 & 29.973 & 0.028 & 8.2 & 226.6 & 35.4 & 120.8 \\
            \noalign{\smallskip}
            \hline
            \noalign{\smallskip}
            1999\,CE119 & 39.583 & 0.274 & 1.473 & 352.711 & 34.967 & 171.553 \\
            2001\,FU172 & 39.636 & 0.272 & 24.694 & 30.943 & 135.196  & 32.448 \\
            \noalign{\smallskip}
            \hline
         \end{array}    
         $$ 
   \end{table}
In our model, we will omit the perturbation of other giant planets like Jupiter or Saturn, in order to reveal the pure effect of CEs brought by Plutinos. Therefore, the model is extremely simplified leaving only the Sun, Neptune and the interacting planetesimals Plutino and Trojan.
This simplification will not significantly impair the reliability of our model, because the secular perturbation from other planets is impossible to directly affect a transient event like CE, whose duration is as short as about 0.1yr. One may argue that the secular perturbation may modify the orbits of Plutino or Trojan in the long term,  whereby CEs are influenced indirectly. But such mechanism is bound to be statistically covered in whole. Only when the secular perturbation persistently and unidirectionally pushes some orbital element of the whole population will it deflect the basic result of the realistic effect by our model, which is very unlikely.
To merely explore the effects of orbital characteristics and also for simplification, all Trojans and Plutinos will be given the mass of Pluto, regardless of their realistic mass. As for the effect of mass, we will develop further analyses in the accompanying paper. Since the Pluto mass is extremely small compared to the Sun or Neptune, we are basically dealing with a restricted 3-body problem (the Sun, Neptune, Trojan) under the perturbation of a 4th body (Plutino).
    
We handle the above model based on the SyMBA package, with a modification to address the CE process and output the crucial information. The CE threshold (denoted by $R_{th}$) is empirically set as $3.5$ Hill radius of Plutino, within which a CE will be identified. The general step-size is set as 0.1 yr, so that the length that Plutino travels in a step (about 0.1 AU) is smaller than the CE threshold. All numerical simulations will be run for $1$Gyr, an order of magnitude of the age of the solar system.
    
We now use CE location and CE time to denote the closest position and the corresponding time during CE. The CE effect will be reflected in the inclination and eccentricity change of Trojan during CE, which is simply the difference between the CE exit and the entry. The CE location, time and effect will be the major roles in the following statistical analyses.
    \subsection{Theoretical evaluation of CE effects}\label{Theo}
    \subsubsection{The Gaussian formula}
    Before presenting the simulation results, we try to look into the scenario of close encounter theoretically and find out what the key elements contributing to the CE effects are.
    
    \begin{figure}[htbp]
    \centering
    \includegraphics[width=0.9\hsize]{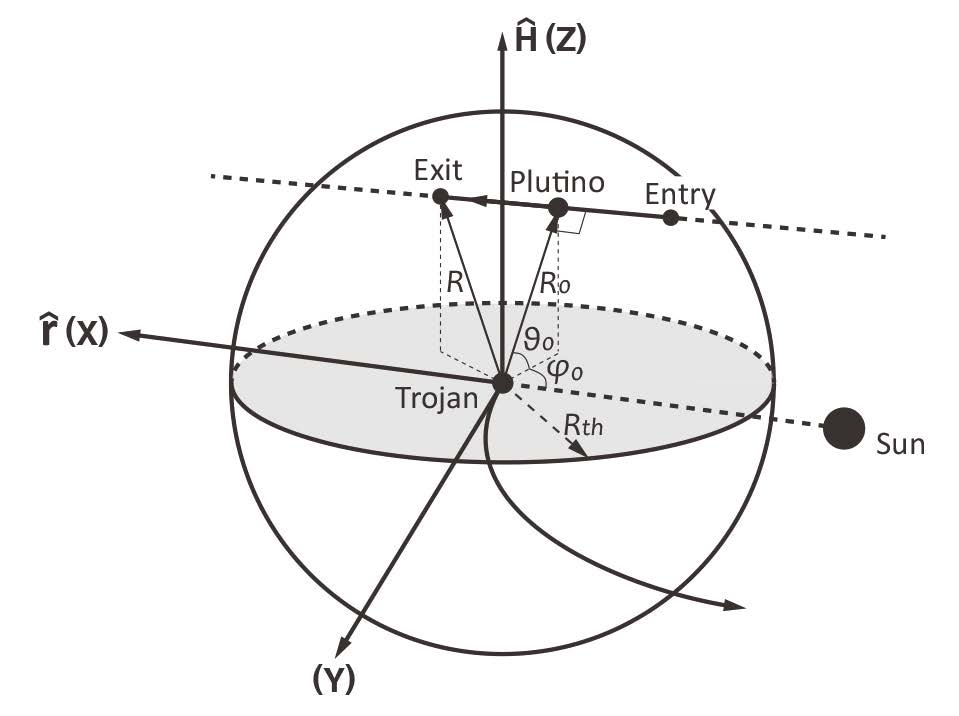}
      \caption{The reference frame applied in the theoretical calculation of CE effects, with exaggerated spatial scale. The reference frame comoves with Trojan, with the Sun on the negative $x$-axis and the velocity of Trojan directed along the $y$-axis. The CE region is a 3-D sphere centering on the Trojan with a radius of 3.5 Hill radius of Plutino. The trajectory of Plutino passing by can be simplified to be rectilinear.}
         \label{illu1}
    \end{figure}   
    
As shown in Fig.\,\ref{illu1}, the reference frame adopted in this section originates at Trojan, with $z$-axis normal to the orbital plane of Trojan and the negative $x$-axis pointing to the Sun. This instantaneous reference frame co-moves with Trojan and is suitable for the Gaussian perturbation equation, where the part concerned with inclination can be expressed as \citep[e.g. ][]{murray_dermott_2000}
    \begin{equation} 
    \label{Gaussian}
    \frac{dI}{dt} = \frac{r\widetilde{N}\cos{(\omega+f)}}{h},
    \end{equation}
which only contains the normal component of the disturbing acceleration $\widetilde{N}$. Here $I$, $\omega$, $f$ corresponds to the inclination, argument of pericentre, and true anomaly of Trojan respectively, and $h$ is the constant associated with the angular momentum of Trojan.
        
The total inclination change during one CE can be written as
    \begin{equation}
    \Delta I = \int_{t_b}^{t_e} \frac{r\widetilde{N}\cos{(\omega+f)}}{h} dt,
    \end{equation}
where $t_b$ and $t_e$ are the time that the integration begins and ends respectively. Here $r$, $\cos{(\omega+f)}$ and $h$ can be considered to be independent of time since the spatial scale of CE is extremely small compared to the orbit circumference of Trojan. Specify $\widetilde{N}$ and we can get
    \begin{equation}
    \label{eq11}
    \Delta I = \frac{\mu r\cos{(\omega+f)}}{h} \int_{t_b}^{t_e} \frac{\bm{R}\bm{\cdot}\bm{H}}{R^3} dt,
    \end{equation}
where $\bm{R}$ is the relative position vector, $\bm{H}$ is the unit normal direction vector of the orbital plane of Trojan and $\mu=Gm_P.$\footnote{From now on, subscripts $T$, $P$ and $S$ correspond to Trojan, Plutino and the Sun respectively. Variables without subscript pertain to Trojan by default.} Due to the low mass of planetesimals and the small scale of the CE region, the motion of Plutino in the vicinity of Trojan is assumed to be rectilinear (see Fig.\,\ref{illu1}). Such approximation has been often adopted in previous works, e.g. \citet{1967JGR....72.2429W}. Hence rewrite Eq.\,(\ref{eq11}) as
    \begin{equation}
    \Delta I = \frac{\mu r\cos{(\omega+f)}}{h} \int_{t_b}^{t_e} \frac{\left(\bm{R_0}+\bm{V_0}t\right)\bm{\cdot}\bm{H}_0}{\left({R_0}^2+{V_0}^2t^2\right)^{3/2}} dt,
    \end{equation}
where for simplicity, the relative velocity $\bm{V}$ is considered to be time-independent during CE and can be substituted by $\bm{V_0}$, which is the relative velocity vector at the closest position\footnote{In this section, subscript $0$ refers to the variable at the closest CE position.}. $\bm{R_0}$ and $\bm{H}_0$ are the relative position vector and the unit direction vector at the closest position respectively. Given the rectilinear trajectory, we can integrate the perturbation from infinity, which means $t_b\to-\infty$ and $t_e\to\infty$. This leads to the total change as
    \begin{equation}
    \label{theoResult}
    \Delta I =  \frac{2\mu r\cos{(\omega+f)}}{h}\frac{\sin{\theta_0}}{V_0}~\frac{1}{R_0},
    \end{equation}
where $\sin{\theta_0}=\bm{R_0}\bm{\cdot}\bm{H}/R_0$. Now if we write $r=a(1-e^2)/\left(1+e\cos f\right)$ and $h=\sqrt{\mu_Ta(1-e^2)}$, where $\mu_T\equiv G\left(m_{T}+m_{S}\right)$, and consider  that $a$, $e$ varies in a narrow range during the evolution of Trojan, the theoretical expression for $\Delta I$ will merely contain $\omega$, $f$, $\theta_0$, $V_0$, $R_0$ as independent variables.
    
We can simplify the result even more by $V_0=\left| \bm{V_P} - \bm{V_T} \right|$, where $\bm{V_P}$ is the velocity vector of Plutino at this moment as well as $\bm{V_T}$ is that of Trojan. Recall the \textit{vis viva} equation
    \begin{equation}
    V_{*}^2=\mu_{*}\left(\frac{2}{r_{*}}-\frac{1}{a_{*}}\right).
    \end{equation}
In our model, $r_{*}\approx a_T$ since all CEs take place close to the orbit of Trojan. Thus
    \begin{equation}
    \label{VPVT}
    \begin{aligned}
    {V_P}^2 &= \mu_P\left(\frac{2}{a_T}-\frac{1}{a_P}\right),\\
    {V_T}^2 &= \frac{\mu_T}{a_T},
    \end{aligned}
    \end{equation}
where $\mu_P\equiv G(m_P+m_S)$.
The relative velocity can be then represented as
    \begin{equation}
    {V_0}^2= {V_P}^2 + {V_T}^2 - 2 V_P V_T \cos{\alpha_v},
    \end{equation}
where $\alpha_v$ is the angle between $\bm{V_T}$ and  $\bm{V_P}$.
Considering both the mass of Trojan and Plutino are fairly small compared to the Sun, we have $\mu_T\approx\mu_P\approx \mu_S\equiv Gm_S$, which leads to
     \begin{equation}
    {V_0}^2= \mu_S \left[ \frac{3}{a_T} - \frac{1}{a_P} - 2 \sqrt{\left( \frac{2}{a_T}-\frac{1}{a_P}\right)\frac{1}{a_T}} \cos{\alpha_v} \right].
    \end{equation}
Hence the inclination change during one CE can be finally written as
    \begin{equation}
    \label{theoResult2}
    \Delta I =  \frac{m_P}{m_S} \frac{a_T}{R_{th}} \sqrt{2\left(1-{e_T}^2\right)} ~\frac{\rho\left(e_T,f_T,\omega_T\right)\sin{\theta_0}}{\sqrt{A - B\cos{\alpha_v}}}~\frac{1}{\gamma_R},
    \end{equation}
where
    \begin{equation}
    	    \gamma_R \equiv\frac{R_0}{R_{th}}, ~~~ A\equiv\frac{1}{2}\left(3 - \frac{a_T}{a_P}\right), ~~~ B\equiv\sqrt{2-\frac{a_T}{a_P}},
    \end{equation}
and
    \begin{equation}
        	    \rho\left(e_T,f_T,\omega_T\right)\equiv\frac{\cos{\widetilde{\lambda_T}}}{1+e_T\cos{f_T}}, ~~~ \widetilde{\lambda_T}\equiv\omega_T+f_T.
    \end{equation}
Here we use $R_{th}$, namely the CE threshold, to normalize $R_0$. The rightmost term $1/\gamma_R$ may need a little bit of modification in practice, which will be discussed later.
    
Expression (\ref{theoResult2}) only consists of a few variables, namely the relative distance $R_0$ and four arguments $\widetilde{\lambda_T}$, $f_T$, $\alpha_v$ and $\theta_0$. These variables can be further divided into two classes, $R_0$ and $\theta_0$ as microscopical since they are defined at the scale of CE, and $\widetilde{\lambda_T}$, $f_T$ and $\alpha_v$ as macroscopical since they are concerned with the orbits of planetesimals. Technically, $R_0$ and $\theta_0$ are determined by $\widetilde{\lambda_T}$, $f_T$, $\alpha_v$ and other macroscopical variables concerning the positions and velocities of planetesimals when CE happens, whereas they can yet be treated to be independent of each other statistically for the entirely different scale they reside in. Therefore we can basically hold that $\Delta I = \Delta I (R_0, \theta_0; \widetilde{\lambda_T}, f_T, \alpha_v)$. 
    
In addition, the term $\cos{\widetilde{\lambda_T}}$ in Eq.\,(\ref{theoResult2}) implies that the disturbing force will have a stronger effect at the ascending or descending node. We can further infer that the relative distance $R_0$ will strongly affect the magnitude of $\Delta I$, while $\theta_0$ and $\widetilde{\lambda}_T$ will directly determine whether the effect is positive or negative.
    
\begin{figure}[htbp]
    \centering
    \includegraphics[width=\hsize]{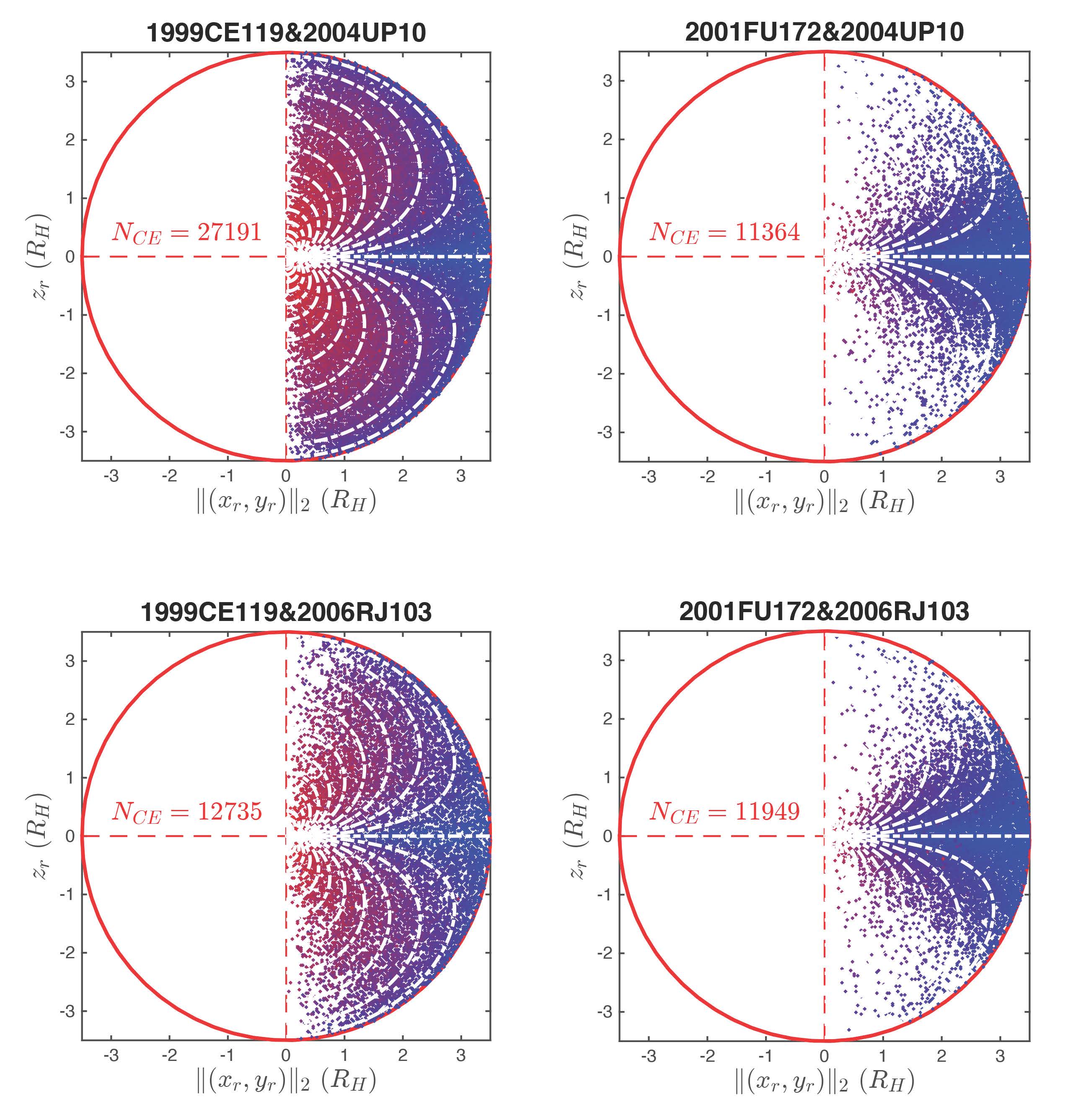}
      \caption{The vertical location distribution of CEs between the Plutinos \& Trojans listed in Table \ref{orbel}. The highlighted boundary indicates the CE sphere. The Trojan lies in the centre, while dots represents the closest relative positions of Plutino during a CE, in a reference frame sketched in Fig.\,\ref{illu1}. Concretely, the ordinate denotes the $z$ component of the relative position vector from Trojan to Plutino as well as the abscissa denotes the modulus of the remaining two components. All lengths are based on the Hill radius of Plutino. In each pair, Plutino clones are introduced to enrich the details (see text in Sect. \ref{CElc}). The color indicates the degree of absolute value of $\Delta I$ in each CE. The more reddish, the higher the effect. The white dashed lines are the contour of $\Delta I=\Delta I(R_0,\theta_0)$ based on Eq.\,(\ref{theoResult2}), with other variables fixed.}
         \label{Stat_CExDstb4}
    \end{figure}

Hereto we can verify the theoretical formula based on (\ref{theoResult}) by comparing it to the simulation results. In Fig.\,\ref{Stat_CExDstb4}, we plot the locations of CEs within the CE region with the color representing the degree of CE effect, while the partial correlation $\Delta I=\Delta I(R_0, \theta_0)$ can be reflected using the contour line in the same figure, which fits the simulation data fairly well.
    
The CE effect on the eccentricity of Trojan can be evaluated in the same way. Gaussian perturbation equation gives \citep{murray_dermott_2000}
    \begin{equation}
    \frac{de}{dt} = \frac{h}{\mu_T} \left[ \widetilde{R} \sin f + \widetilde{T}\left( \cos f + \cos E \right) \right].
    \end{equation}
By several simple derivations, we have
    \begin{equation}
    \label{theoResulte}
    \Delta e = \frac{m_P}{m_S}\frac{a_T}{R_{th}}\sqrt{2\left(1-{e_T}^2\right)}~\frac{\varsigma\left(e_T,f_T,\varphi_0\right)\cos{\theta_0}}{\sqrt{A-B\cos{\alpha_v}}}~\frac{1}{\gamma_R}, 
    \end{equation}
    where
    \begin{equation}
    \begin{aligned}
    & \varsigma(e_T,f_T,\varphi_0)=\sin{(f_T+\varphi_0)}+\cos E_T\sin{\varphi_0}, \\
    & \cos E_T = \frac{e_T+\cos f_T}{1+e_T\cos f_T},
    \end{aligned}
    \end{equation}
    and
    \begin{equation}
    \label{Eq:Phi0Def}
    \begin{aligned}
    & R_0\cos\theta_0\cos\varphi_0 =\bm{R_0}\bm{\cdot}\bm{R}, \\
    & R_0\cos\theta_0\sin\varphi_0 =\bm{R_0}\bm{\cdot}\bm{T}.
    \end{aligned}
    \end{equation}
    Eq.\,(\ref{Eq:Phi0Def}) defines $\varphi_0$, where $\bm{R_0}$ and $\theta_0$ have been previously defined while $\bm{R}$ and $\bm{T}$ are the unit radial and transversal vector in the reference frame illustrated in Fig.\,\ref{illu1} respectively. Here we take the similar approximations as above and the expressions for $\Delta I$ and $\Delta e$ are basically similar. 

\subsubsection{\textbf{Compared with \"Opik's formula}}
  
Although the above theoretical evaluation presents a clear physical description of CE via Gaussian perturbation equations, we are aware that the classical \"Opik's formula \citep{1976iecg.book.....O} is more regularly applied in previous works  \citep[e.g. ][]{1988Icar...75....1G, 1990CeMDA..49..111C, 2015CeMDA.123..151V} to handle similar cases. We now compare our above formula (hereinafter Gaussian formula) with \"Opik's formula.
    
\"Opik's formula is able to predict the entire set of post-encounter orbital elements, based on the pre-encounter elements of the two interacting planetesimals. While in this work, we only care about the CE effects, namely the change of inclination and eccentricity. To output the outcome orbits using \"Opik's formula and then seek the minor differences seems rather complicated and overqualified. In contrast, the Gaussian formula is certainly more direct and convenient. One more difference lies in the input parameters. As shown in Eqs.\,(\ref{theoResult2}) and (\ref{theoResulte}), the input variables of Gaussian formula are mainly the CE parameters that denote the locations and orientations of CEs. Such information happens to be our main concern. Later in this section we will discuss these parameters one by one and derive the corresponding distributions for numerous CEs, and finally lead to the distribution of CE effects. This route gives us an explicit pattern about the relationship between CE parameters and dynamic effects, based on the physical connections offered by Gaussian formula. For \"Opik's formula, the distributions of pre-encounter orbital elements are hard to obtain and also not very helpful in the understanding of CEs with regard to our concern.
     
    \begin{figure}[htbp]
    \centering
    \includegraphics[width=\hsize]{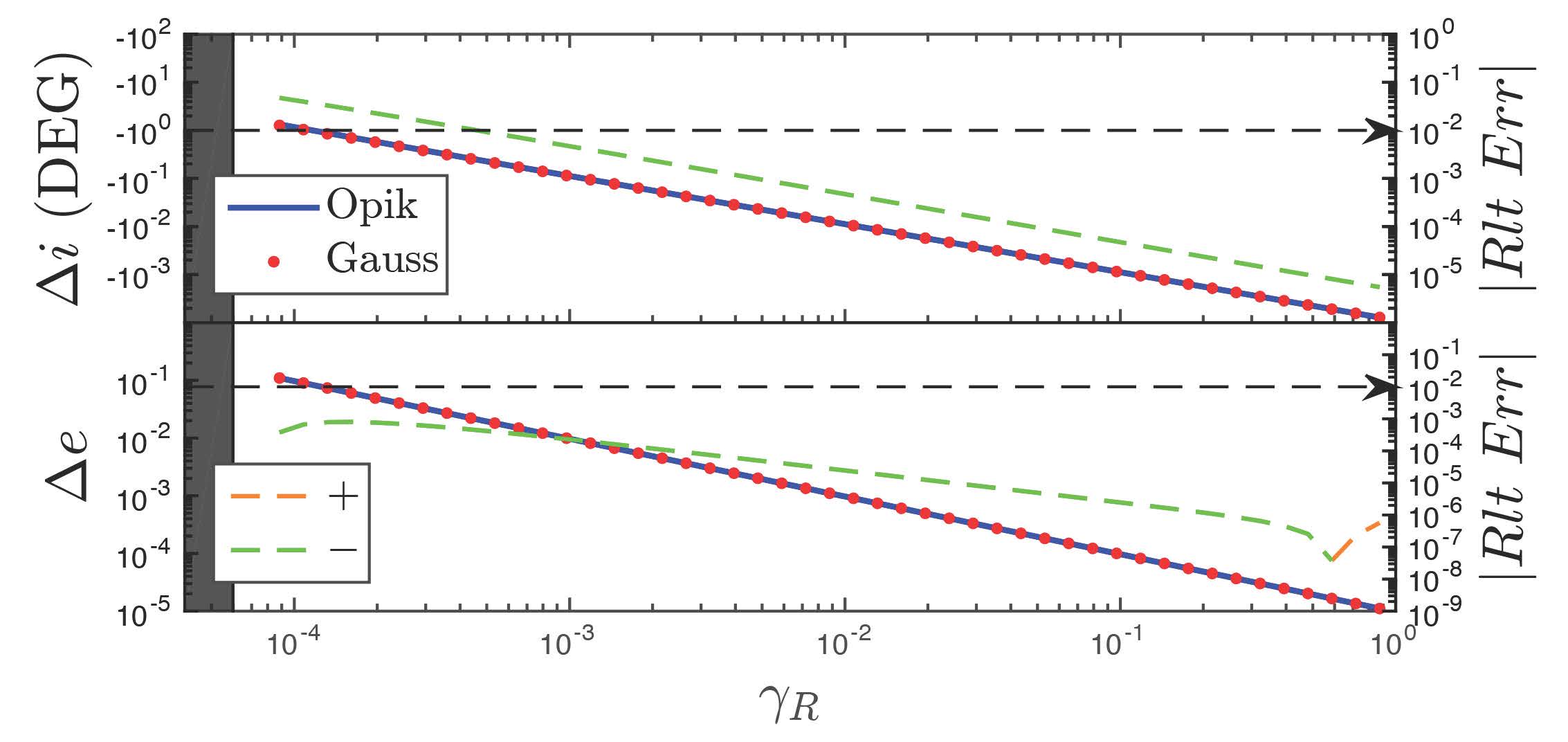}
      \caption{The CE effects ($\Delta I$ and $\Delta e$) calculated by \"Opik (Blue curves) and Gaussian (Red dots) formulas as the CE distance varies. The relative error between the two, namely $|Rlt~Err|=|(Gaussian-\ddot{O}pik)/\ddot{O}pik|$, is shown on the right axis by the orange (if positive) and green (if negative) dashed curve. The black dashed line denotes the $1\%$ critical error. The region where CE distance is less than the physical radius of the planet (about 1188 km for Pluto) is shadowed. 
      }
         \label{OpikGaussComparison_radius}
    \end{figure}
    
Nevertheless, for the sake of conciseness and understandability, the Gaussian formula introduces the rectilinear approximation, which is after all rough and may damage its performance in certain cases. In order to verify the reliability and practicability of Gaussian formula, we now compare its result with that of \"Opik's formula, which is generally more sophisticated. The model will include a hypothetical planet with Pluto mass, situated at 30AU, moving in a circular orbit, and a particle at 40 AU, with $e=0.25$ to make the CE possible. 

Fig.\,\ref{OpikGaussComparison_radius} juxtaposes the CE effects calculated by both formulas. As the CE distance increases, the CE effects decrease sharply, which is consistent with Eq.\,(\ref{theoResult2}). Generally speaking, the result of Gaussian equation is close to that of \"Opik's formula, with minor underestimation because the rectilinear approximation makes the particle a little bit farther from the planet. The relative error curve further shows that deviation turns greater when $\gamma_R$ is smaller. This tendency is easy to understand for that when a CE is close to the surface of the planet, huge perturbation must bend the trajectory more and depreciate the accuracy of rectilinear approximation.
However, this relative higher central deviation will not impair the practicability of Gaussian formula. As shown in Fig.\,\ref{OpikGaussComparison_radius}, the absolute relative error between Gaussian and \"Opik will turn significant (1\% generally) only when $\gamma_R\lesssim 5\times 10^{-4}$, in which case, as will be later proven in Eq.\,(\ref{Dis2Dstb}), the probability of CE will be as low as $10^{-6}$. Considering that a general pair of planetesimals will merely generate a few thousand CEs in 1Gyr, a CE so close to the surface will rarely happen even for a large group of planetesimals. 
     
In Fig.\,\ref{OpikGaussComparison_radius}, one may notice another feature that the relative error for $\Delta e$ rises abnormally and turns positive in the far right end. This is due to the infinite integration in the derivation of Gaussian formula, which is acceptable when the CE scale is small, but clearly inappropriate when the relative distance is comparable to the length of the orbit.
    
   \begin{figure}[htbp]
    \centering
    \includegraphics[width=\hsize]{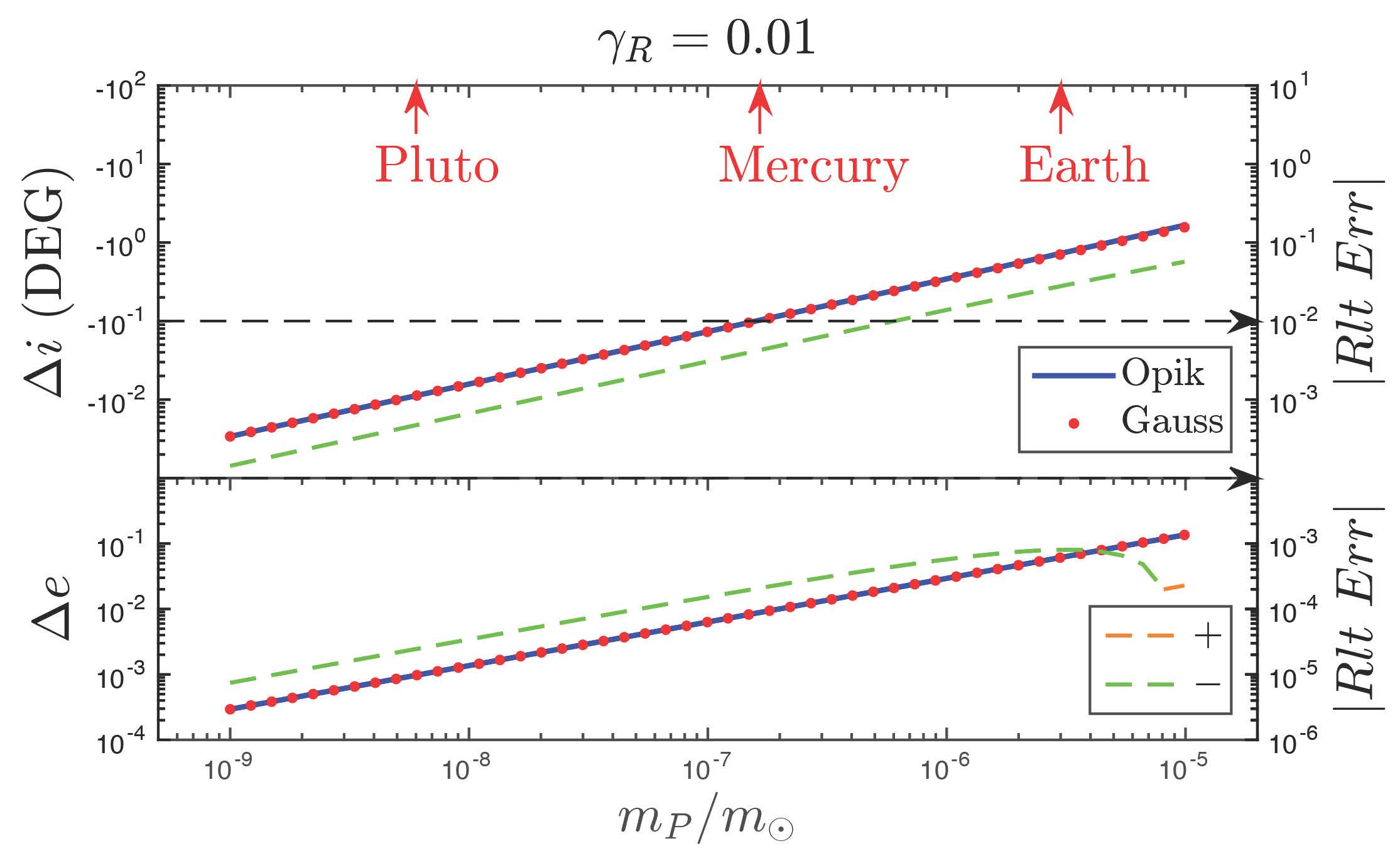}
      \caption{The CE effects calculated by \"Opik and Gaussian formulas as the mass of the planet varies. The detailed legends are the same as Fig.\,\ref{OpikGaussComparison_radius}. The realistic masses of Pluto, Mercury and Earth are annotated on the top for reference.}
      \label{OpikGaussComparison_mass}
    \end{figure}
    
In Fig.\,\ref{OpikGaussComparison_mass}, we test the performance of Gaussian formula with respect to the mass of the planet. $\gamma_R$ is fixed to be 0.01. The mass of the planet ranges from $10^{-9}$ to $10^{-5}$ times the Sun mass, which are the magnitude of Pluto mass and Earth mass respectively. As the mass increases, the CE effect and the relative error both become greater, which is due to the same reason mentioned before. At about $10^{-6}$ times the Sun mass, the relative error of $\Delta I$ exceeds the $1\%$ critical value. This is slightly higher than the mass of Mercury. Note that on the bottom panel in Fig.\,\ref{OpikGaussComparison_mass}, the relative error of $\Delta e$ bends abnormally at high mass, which leads to another deficiency of Gaussian equations. According to Eqs.\,(\ref{theoResult2}) and (\ref{theoResulte}), the CE effect will always increase proportionally with the mass of the planet, \emph{ceteris paribus}. However, $\Delta I$ and $\Delta e$ have natural restrictions, i.e. $\pi$ for $I$ and 1 for $e$. This implies that when the mass of the planet is high, the result of Gaussian formula may not be reasonable. 
    
In general, Gaussian formula is applicable to CEs that are not very close (farther than 0.0005 CE threshold), and disturbed by bodies comparable to asteroids in mass. Our problem meets this requirement precisely.
    
\subsection{Distribution of CE location}\label{CElc}
As shown in Eq.\,(\ref{theoResult2}) and Eq.\,(\ref{theoResulte}), the CE effects are significantly influenced by $\theta_0$, $\varphi_0$ and $R_0$, which actually determine the location of CE in the CE sphere, as illustrated by Fig.\,\ref{illu1}. Therefore, in this section, we will focus on the distribution of these variables, while for convenience, the distributions of $\theta_0$, $\varphi_0$ and $R_0$ will be referred to as the azimuthal and radial distribution respectively.

Intended for more CE data to reveal a more clear and accurate distribution, 40 clones of the corresponding Plutino in each pair are introduced with arbitrary value of orbital elements. Concretely, we first carry out a simple simulation on the model including the Sun, Neptune and the realistic Plutino with Pluto mass, and record the upper and lower limits of the orbital elements of the Plutino during evolution, within which the orbital elements of clones are arbitrarily picked. Note that the angles like $M$, $\omega$ and $\Omega$ are picked within $0$ and $2\pi$ in radians straightforwardly. 
    
    \subsubsection{The azimuthal symmetry of vertical CE location}
    \label{CEAzmSym}
    
    \begin{figure}[htbp]
    \centering
    \includegraphics[width=\hsize]{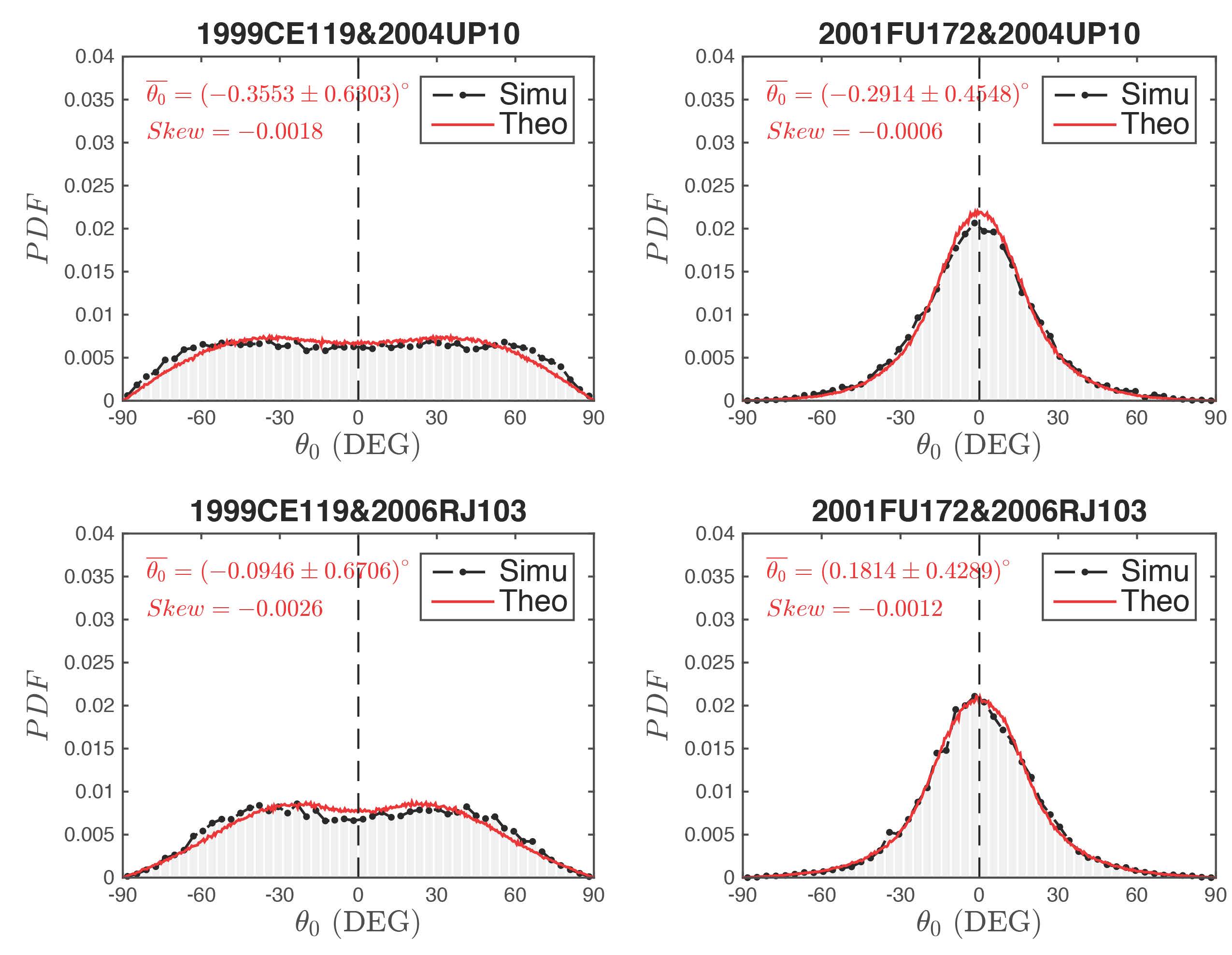}
      \caption{The probability density function of the latitude of CE location $\theta_0$. The dotted curves are from numerical simulations. The positive and negative branches are quite symmetric, reflected by the small skewness annotated on the top. The mean value, and its upper and lower limits at the 5\% significance level are annotated as well. The expected $\overline{\theta_0}=0$, denoted by a dashed line, is within the margin of error for each pair.  The highlighted curves are the theoretical predictions (see text in Sect. \ref{Sec:Theta0Theo}).}
         \label{Symba_azmThetaDstb}
    \end{figure}
    
We have already shown the locations of CEs in Fig.\,\ref{Stat_CExDstb4}, distributing quite symmetrically around $z=0$, which corresponds to the instantaneous orbital plane of Trojan. To clearly reveal the symmetry between positive and negative branches, Fig.\,\ref{Symba_azmThetaDstb} shows the distribution of $\theta_0$ defined in Eq.\,(\ref{theoResult}), which can be alternatively expressed as
    \begin{equation}
    \theta_0 \equiv \arctan{\frac{z_r}{\Vert(x_r,y_r)\Vert_2}},
    \end{equation}
intuitively the latitude of the CE location in the comoving reference frame. $x_r$, $y_r$ and $z_r$ here represent the three components of the spatial vector from Trojan to Plutino.
    
To further verify if the positive and negative branches share the same distribution, we introduce the two-sample Kuiper test \citep{1995sacd.book.....F}. The null hypothesis is the two distributions are identical. At the 5\% significance level, all pairs obey the null hypothesis, which points out a highly symmetric distribution of CE location.

\subsubsection{The theoretical distribution of $\theta_0$}
  \label{Sec:Theta0Theo}
Fig.\,\ref{Symba_azmThetaDstb} also shows that a Plutino or Trojan with high inclination will result in CEs gathering around $\theta_0=0$, i.e. $z=0$ (two panels in the right column). Considering the particular situation during CE, this tendency should result from the orientation bias of the relative velocity vector due to different inclined level of the orbital plane of interacting planetesimals. Intuitively, one can imagine that when the inclination of Plutino is relatively large, the CE location, geometrically the tangent point between the trajectory of Plutino and the spherical surface of CE region centering on Trojan, is unlikely to locate at a high latitude. 
    	
To give a quantitative explanation and thus obtain the theoretical distribution of $\theta_0$, we will start from the theoretical estimations on the alignments of the velocity vectors of Plutino and Trojan, whereby $\theta_0$ can be derived through geometrical relations.
    
\begin{figure}[htbp]
    \centering
    \includegraphics[width=\hsize]{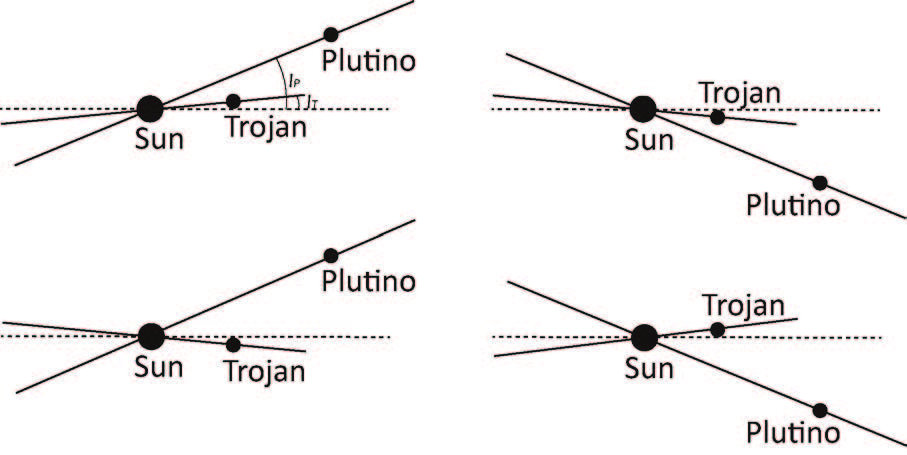}
      \caption{The illustration for the four different relative positions between the orbital planes of Plutino and Trojan, which result in the four branches of the distribution of $\theta_{v_P}$.}
      \label{illu_thetavp}
    \end{figure}

Concretely, in the first place, a von Mises distribution \citep{1995sacd.book.....F} is applied to the approximation of the distribution of $\theta_{v_P}$, defined as
        \begin{equation}
        	\theta_{v_P} \equiv  \arctan \frac{v_{P,z}}{\sqrt{v_{P,x}^2+v_{P,y}^2}},
        \end{equation}
namely the latitude of the velocity vector of Plutino in the comoving coordinate. $v_{P,x}$, $v_{P,y}$ and $v_{P,z}$ here represent the three components of the velocity vector of Plutino.
Naturally, the value of $\theta_{v_P}$ is close to the inclination of Plutino. When the inclination of Trojan is further considered, the distribution of $\theta_{v_P}$ will split into four branches, as illustrated in Fig.\,\ref{illu_thetavp}. Therefore the probability density function of $\theta_{v_P}$ can be given as
        \begin{equation}
        \label{PDFthetavp}
        	f(\theta_{v_P}) = \frac{1}{8\pi I_0\left(\kappa_f\right)}\sum_{i=1}^4{\exp{\left[\kappa_f\cos{\left(\theta_{v_P}-\overline{\theta_{v_P}^{\,i}}\right)}\right]}},
        \end{equation}
where $\kappa_f$ is a measure of concentration, $I_0$ is the modified Bessel function of order 0, and 
        \begin{equation}
        	\overline{\theta_{v_P}^{\,i}} = \pm I_P \pm I_T,~~ i=1,2,3,4.
        \end{equation}

Similarly, we define
         \begin{equation}
        	\varphi_{v_P} \equiv \arctan \frac{v_{P,y}}{v_{P,x}},
        \end{equation}
namely the longitude of velocity vector of Plutino.
The theoretical distribution of $\varphi_{v_P}$ can be estimated as the superposition of two von Mises distribution, namely
      \begin{equation}
      \label{PDFphivp}
       	f(\varphi_{v_P}) = \frac{1}{4\pi I_0\left(\kappa_f\right)}\sum_{i=1}^2{\exp{\left[\kappa_f\cos{\left(\varphi_{v_P}-\overline{\varphi_{v_P}^{\,i}}\right)}\right]}}
      \end{equation}
where $\kappa_f$ is the same parameter defined as above and
    \begin{equation}
    \label{meanphivp}
    	\overline{\varphi_{v_P}^{\,i}} = 90 \deg\pm \alpha_{CE},~~ i=1,2.
    \end{equation}
    
    \begin{figure}[htbp]
    \centering
    \includegraphics[width=0.7\hsize]{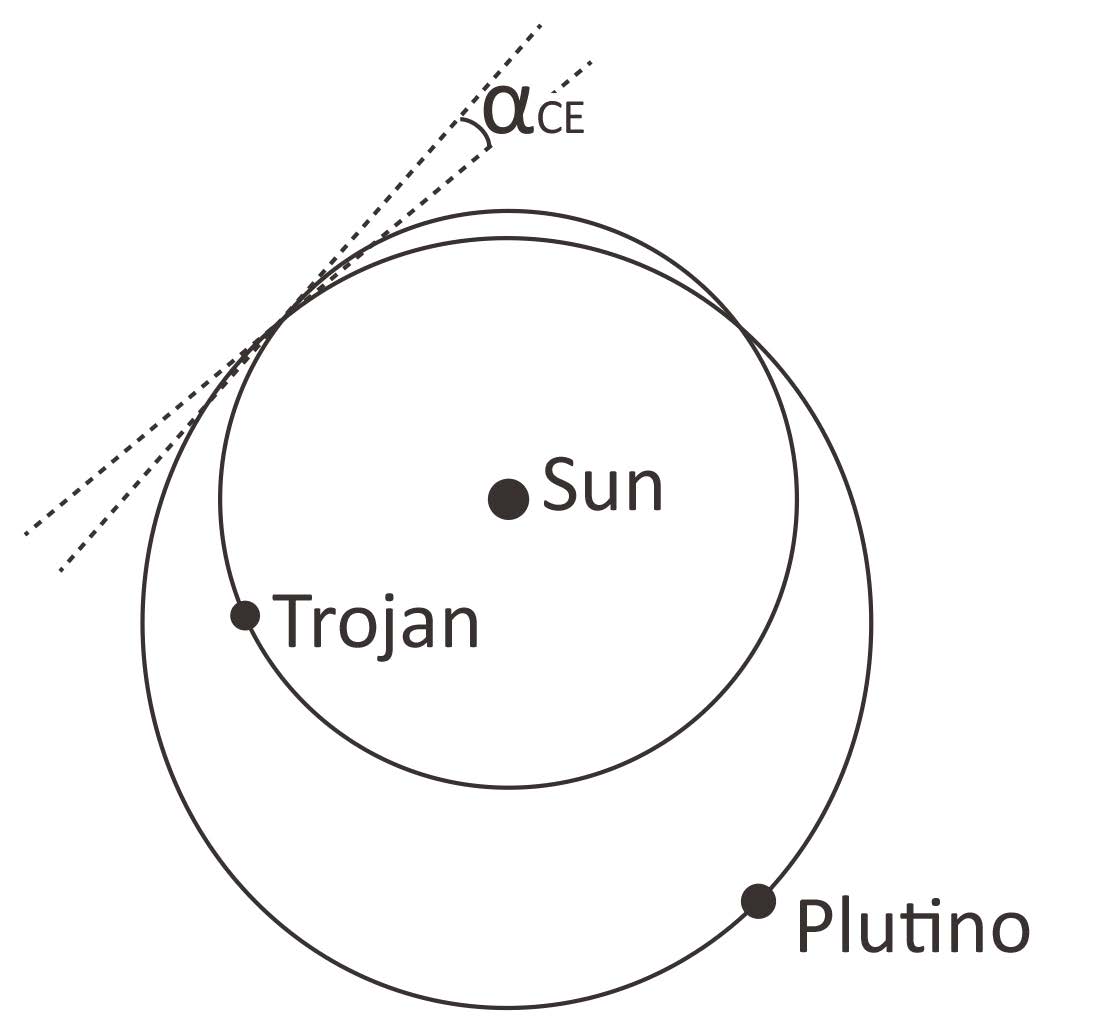}
      \caption{The illustration for $\alpha_{CE}$, namely the angle between the velocity directions of Plutino and Trojan in CE.}
      \label{illu_phivp}
    \end{figure}
    
As illustrated in Fig.\,\ref{illu_phivp}, $\alpha_{CE}$ is the angle between the velocity directions of Plutino and Trojan in CE, namely the tangent separation at the intersection of two trajectories. Since the velocity direction of Trojan is perpendicular to the abscissa in the comoving coordinate (see Fig.\,\ref{illu1}), i.e. the radial direction of Trojan, and the velocity direction of Plutino can be either inward or outward with respect to the trajectory of Trojan, the mean values of $\varphi_{v_P}$ can be roughly linked to $\alpha_{CE}$ as in Eq.\,(\ref{meanphivp}), which then contribute to the two centers in the distribution of $\varphi_{v_P}$. 
    
Note that when the orbits of Plutino and Trojan are inclined differently, the exact definition of $\alpha_{CE}$ is ambiguous, whereas it can still be approximated by hypothetically placing the two orbits on the same plane, since the realistic inclinations of Plutino and Trojan are low for most cases.
    
We now have the theoretical distribution of $\theta_{v_P}$ and $\varphi_{v_P}$ in CE, by which the orientation of relative velocity vector can be obtained geometrically. Concretely, the relative velocity vector can be calculated as
     \begin{equation}
     	\bm{v_r} = (V_P\cos\theta_{v_P}\cos\varphi_{v_P},V_P\cos\theta_{v_P}\sin\varphi_{v_P}-V_T,V_P\sin\theta_{v_P}),
     \end{equation}
where $V_P$ and $V_T$ are the respective modules of the velocities of Plutino and Trojan mentioned in Eq.\,(\ref{VPVT}), which remain basically constant in CE. Therefore, we immediately derive the latitude and longitude of $\bm{v_r}$, respectively
     \begin{equation}
        	\theta_{v_r} \equiv  \arctan \frac{v_{r,z}}{\sqrt{v_{r,x}^2+v_{r,y}^2}},~~~ \varphi_{v_r} \equiv  \arctan \frac{v_{r,y}}{v_{r,x}},
     \end{equation}
where $v_{r,x}$, $v_{r,y}$ and $v_{r,z}$ represent the three components of the relative velocity vector.
 
    \begin{figure}[htbp]
    \centering
    \includegraphics[width=0.75\hsize]{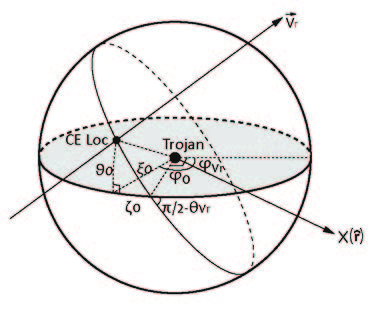}
      \caption{The illustration for the relation between the orientation of relative velocity vector and the CE location. The relatively velocity vector is assumed to lie in the paper surface for simplicity, but without loss of generality. }
         \label{illu_theta}
    \end{figure}   
    
Now with the relative velocity vector determined, we can finally trace the CE picture and solve $\theta_0$.  As illustrated in Fig.\,\ref{illu_theta}, given the specific orientation of $\bm{v_r}$, the CE location will reside at the orthodrome perpendicular to $\bm{v_r}$ on the sphere of CE region. To further determine the CE location, we require an auxiliary variable $\xi_0$, which measure the distance between the CE location and the equatorial plane (Trojan orbital plane). Consequently, with the help of spherical trigonometry,  we have
    \begin{equation}
    	\sin\theta_0  = \cos\theta_{v_r}\sin\xi_0.
    \end{equation}
    
Overall, $\theta_0$ obeys a joint distribution of $\theta_{v_r}$ and $\xi_0$. The distribution of $\theta_{v_r}$ can be derived from the distributions of $\theta_{v_P}$ and $\varphi_{v_P}$, which are given in Eq.\,(\ref{PDFthetavp}) and Eq.\,(\ref{PDFphivp}) respecitively. As for $\xi_0$, in consideration of the microscopical randomness of the relative positions when Plutino flies past Trojan, it can be simply treated as uniform. 
    
Note that for all the von Mises distributions mentioned before, we use the identical parameter $\kappa_f$, making it the only artificial parameter in this derivation. Naturally, $\kappa_f$ involves the dispersion of the velocity orientation brought by randomness, which is not related to the orbital characteristics and should be constant. Nevertheless, due to the improper handling of $\alpha_{CE}$ when Plutino is highly inclined early in this section, we have to introduce an extra term regarding Plutino inclination in the determination of $\kappa_f$. Therefore, the expression of $\kappa_f$ can be empirically derived as 
    \begin{equation}
        \left(\frac{1}{\sqrt{\kappa_f}}\right)/\deg = {\left(I_P/\deg\right)}^{0.6}+3.5.
    \end{equation}
The left side is analogous to the standard deviation $\sigma$ in the normal distribution. We shall later show the low dependency of our analytical model on $\kappa_f$ in the accompanying paper.
    
In practice, due to the complex derivations above, to solve the explicit expression of the distribution of $\theta_0$ will be troublesome. Alternatively, we can obtain $\theta_0$ by generating a large sample of $\theta_{v_P}$ and $\varphi_{v_P}$ distributing as defined by Eqs.\,(\ref{PDFthetavp}) and (\ref{PDFphivp}), and then handle the data statistically. In Fig.\,\ref{Symba_azmThetaDstb}, the theoretical distribution of $\theta_0$ is highlighted, adequately reflecting the statistical characteristics of the numerical result.
        
 \subsubsection{The horizontal distribution of CE location}
    
    \begin{figure}[htbp]
    \centering
    \includegraphics[width=\hsize]{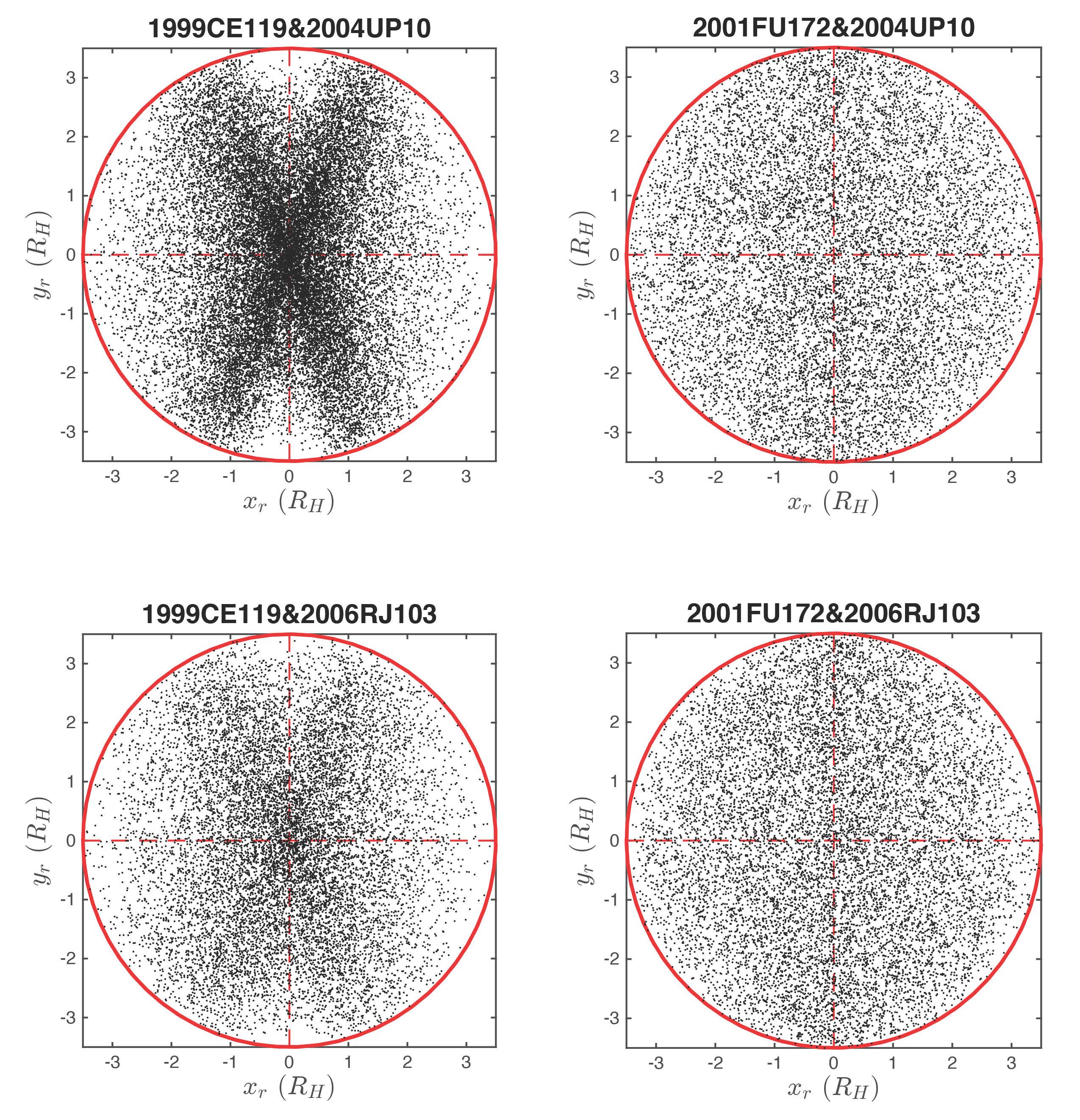}
      \caption{The horizontal location distribution of CEs between the Plutinos \& Trojans listed in Table \ref{orbel}, which is similar to Fig.\,\ref{Stat_CExDstb4}. The prominent ``X'' pattern fades away as inclinations of planetesimals increase, which is a phenomenon concerning the specific orientation of orbits when CEs happen.}
         \label{Stat_CExDstb4xy}
    \end{figure}
    
Besides the vertical distribution of CEs along the $z$-axis, the horizontal distribution, namely how the projections of CE locations distribute on the $(x,y)$ plane also matters, which is thus shown in Fig.\,\ref{Stat_CExDstb4xy}. The abscissa here indicates the $x$ component of the relative position vector from Trojan to Plutino in the reference frame introduced early while the ordinate corresponds to the $y$ component. Therefore the figure actually depicts the projections of CE locations on the orbital plane of Trojan. Note that $y=0$ here is coincident with the heliocentric vector of Trojan. 
A pronounced ``X'' pattern resides within the CE threshold in the top left plot, whereas the pattern gradually fades away as the inclinations of the two interacting planetesimals increase, till a completely uniform distribution at a very high inclination. According to Fig.\,\ref{illu_phivp}, one could picture that, given an appropriate eccentricity, the Plutino has two chances to cross the orbit of Trojan in one cycle, once inward and the other outward, thus creating the two strokes of figure-X. 
However, as the orbits become inclined, the crosses take place in all directions. Though a pattern may be observed in one specific direction, the projection tends to mix up the patterns and a uniform distribution appears in the end. 
    
\begin{figure}[htbp]
    \centering
    \includegraphics[width=\hsize]{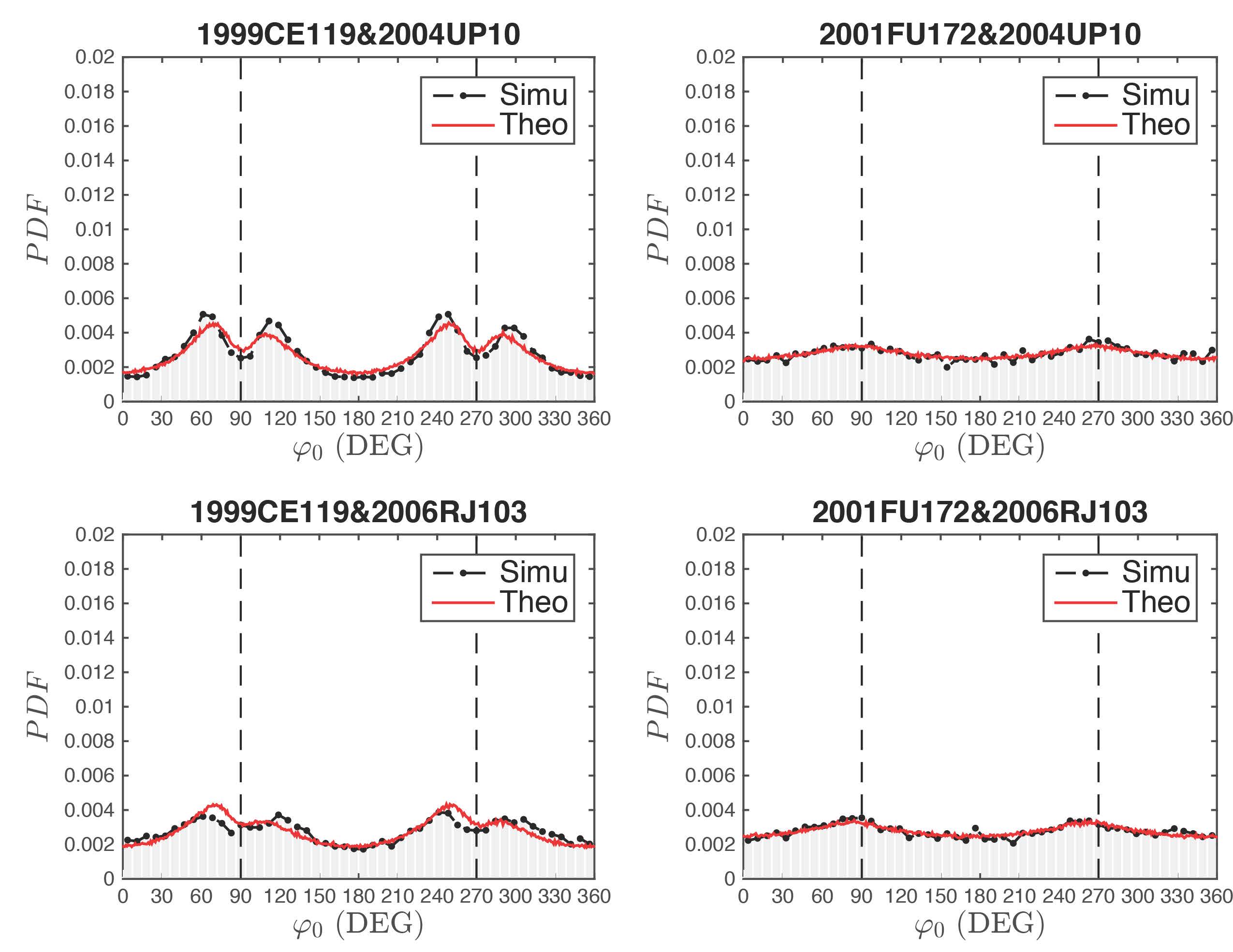}
    \caption{The probability density function of the longitude of CE location $\varphi_0$. The dashed lines indicate $\varphi_0=90\deg$ and $\varphi_0=270\deg$, which coincide with the direction of the orbit of Trojan. The dotted curves are the numerical results and the highlighted curves are the theoretical predictions.}
    \label{Symba_azmPhiDstb}
    \end{figure}

Statiscally, Fig.\,\ref{Symba_azmPhiDstb} presents the distribution of $\varphi_0$, defined in Eq. (\ref{Eq:Phi0Def}), which can be alternatively expressed as
    \begin{equation}
    \varphi_0 \equiv \arctan{\frac{y_r}{x_r}}.
    \end{equation}
One can find that the peaks, corresponding to figure-X in Fig.\,\ref{Stat_CExDstb4xy}, fade away clearly as the inclination of planetesimals increases.
Meanwhile, as illustrated by Fig.\,\ref{illu_theta}, $\varphi_0$ can be analytically obtained through
    \begin{equation}
         \varphi_0 = \varphi_{v_r} - \left(\frac{\pi}{2}+\zeta_0\right),
    \end{equation}
where $\zeta_0$ is the horizontal component of $\xi_0$ and geometrically
    \begin{equation}
    	\tan\zeta_0 = \sin\theta_{v_r} \tan\xi_0.
    \end{equation}
The resulting distributions of $\varphi_0$ are highlighted in Fig.\,\ref{Symba_azmPhiDstb}.
    
    \subsubsection{The radial distribution of CE location}
    
\begin{figure}[htbp]
    \centering
    \includegraphics[width=\hsize]{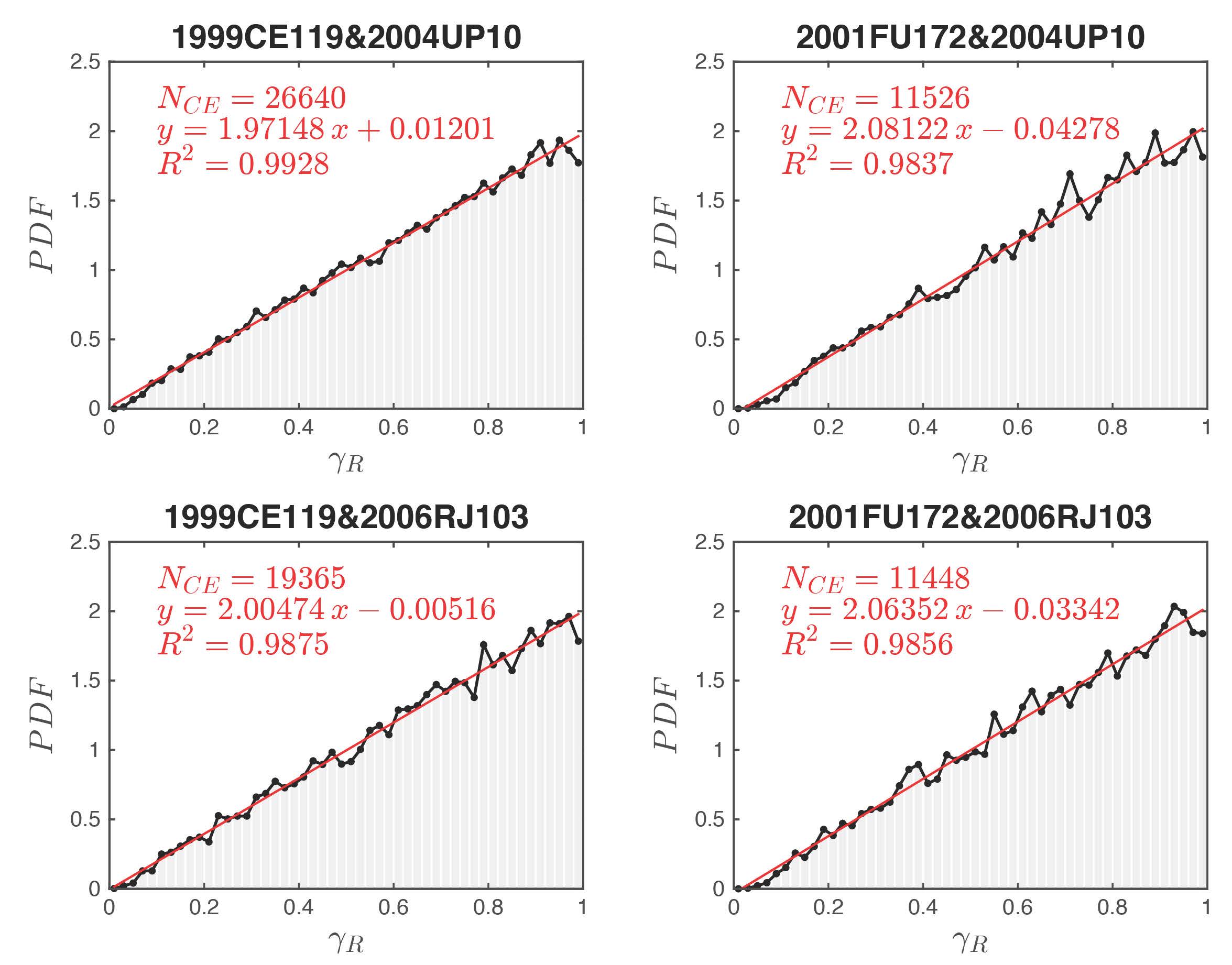}
      \caption{The probability density function of the minimum CE distance $\gamma_R$. The highlighted line depicts the result of linear fitting, which possesses similar fitting parameters for different interacting planetesimals.} 
         \label{Stat_DisDstb4}
    \end{figure}

We are also interested in the radial distribution of the CE locations, namely the distribution of the minimum distances between two planetesimals in CEs. In Fig.\,\ref{Stat_DisDstb4}, the distribution of the minimum distance $\gamma_R$ is presented, turning out to be a linear distribution with a high correlation coefficient. Though the interacting planetesimals differ from one pair to another, the fitting parameters basically remains the same. 
    
In fact, the dependence of CE probability on the minimum distance has been analytically explored in many precedent works \citep{1951PRIA...54..165O, 1967JGR....72.2429W}, either through strict geometrical derivations, or in a classical ``cross section'' view. Therefore, here we directly mark down the correlation in a normalized way, since we are solely concerned with the distribution within the CE sphere. The probability of a CE locating within $R$ will be
    \begin{equation}
    \label{Dis2Dstb}
    P(\widetilde{\gamma_R}<\gamma_R)={\gamma_R}^2,
    \end{equation}
where the variables associated with $R$ are replaced by $\gamma_R$, after being normalized by the radius of CE sphere $R_{th}$. Thereupon the probability density function of $\gamma_R$ is
    \begin{equation}
    \label{DisDeltaDstb}
    f(\gamma_R)=2\gamma_R,
    \end{equation}
coinciding with the fitting parameters in Fig.\,\ref{Stat_DisDstb4} fairly well.

\subsection{Distribution of CE time}
\label{SectimeDstb}

We wonder if the CEs would concentrate on particular time and be absent in other period, worrying about a temporary swarm of CEs causing a strong interaction. Consequently we need to examine the distribution of CE time. Recall that previously we apply the model including the Sun, Neptune and the massless Trojan, where 40 clones of Plutino are introduced for more CE data. Nonetheless, the data concerning clones are not suitable here, since the dynamic instability of clones will always lead to a prominent drop of the quantity of CEs over time and thus contaminate      the intrinsic distribution of CE time. Hence, here we will introduce only one Plutino into the model, with realistic initial value of orbital elements. 
    
     \begin{figure}[htbp]
    \centering
    \includegraphics[width=\hsize]{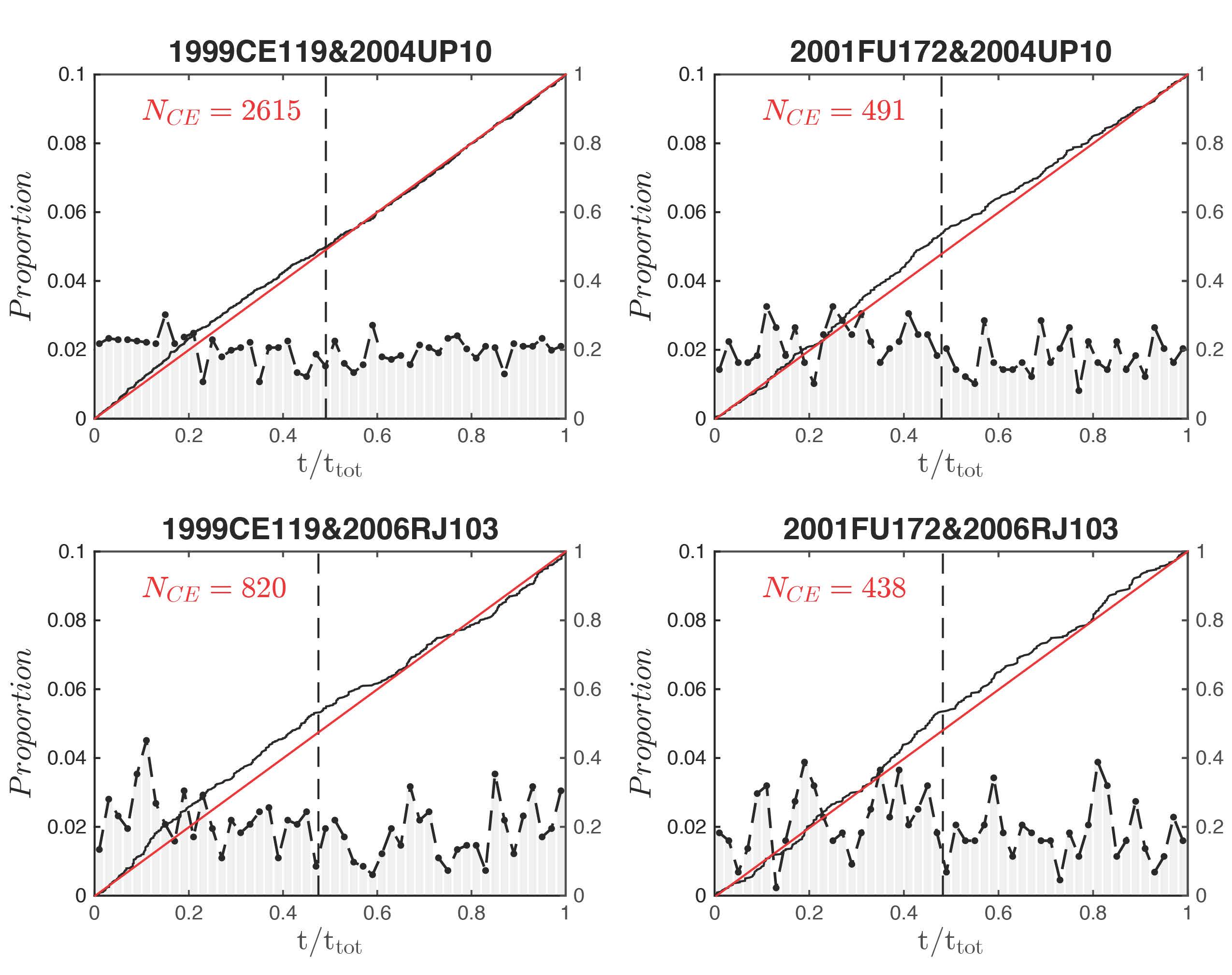}
      \caption{The proportion of number of CEs in each time interval on left ordinate, as well as the proportion of number of CEs before one particular time on right ordinate. The abscissa corresponds to the current time of CE over the total evolution time. The vertical dashed line depicts the mean value of CE time, which is quite close to half the total time. The highlighted line corresponds to the uniform cumulative distribution function.}
         \label{Symba_TimeDstb}
    \end{figure}    
    
Fig.\,\ref{Symba_TimeDstb} shows the frequency histograms of CE time, as well as the empirical cumulative distribution function (ECDF) \citep{van2000asymptotic} which indicates the portion of CEs before one particular time. The highlighted line, which is the cumulative distribution function (CDF) of a uniform distribution, is of course linear. Thus by comparison, the distribution of CE time turns out to be quite uniform. In general, we may consider the CE to take place evenly during evolution, unlikely to cause a strong interaction within a particular period of time.
    
Here we introduce the one-sample K-S test \citep{eadie1971statistical} to check the uniformity. The null hypothesis is that the sample is drawn from the uniform distribution. 
 
At the 5\% significance level, all pairs obey the null hypothesis except Plutino 1999 CE119 and Trojan 2006 RJ103, which we attribute to the lack of sample points. After all, the K-S method is a rigorous test that technically ensures the compliance of the sample with the reference distribution.

\subsection{Distribution of CE effects}
\label{diDstb}
\subsubsection{The numerical distribution of $\Delta I$}
\begin{figure}[htbp]
    \centering
    \includegraphics[width=\hsize]{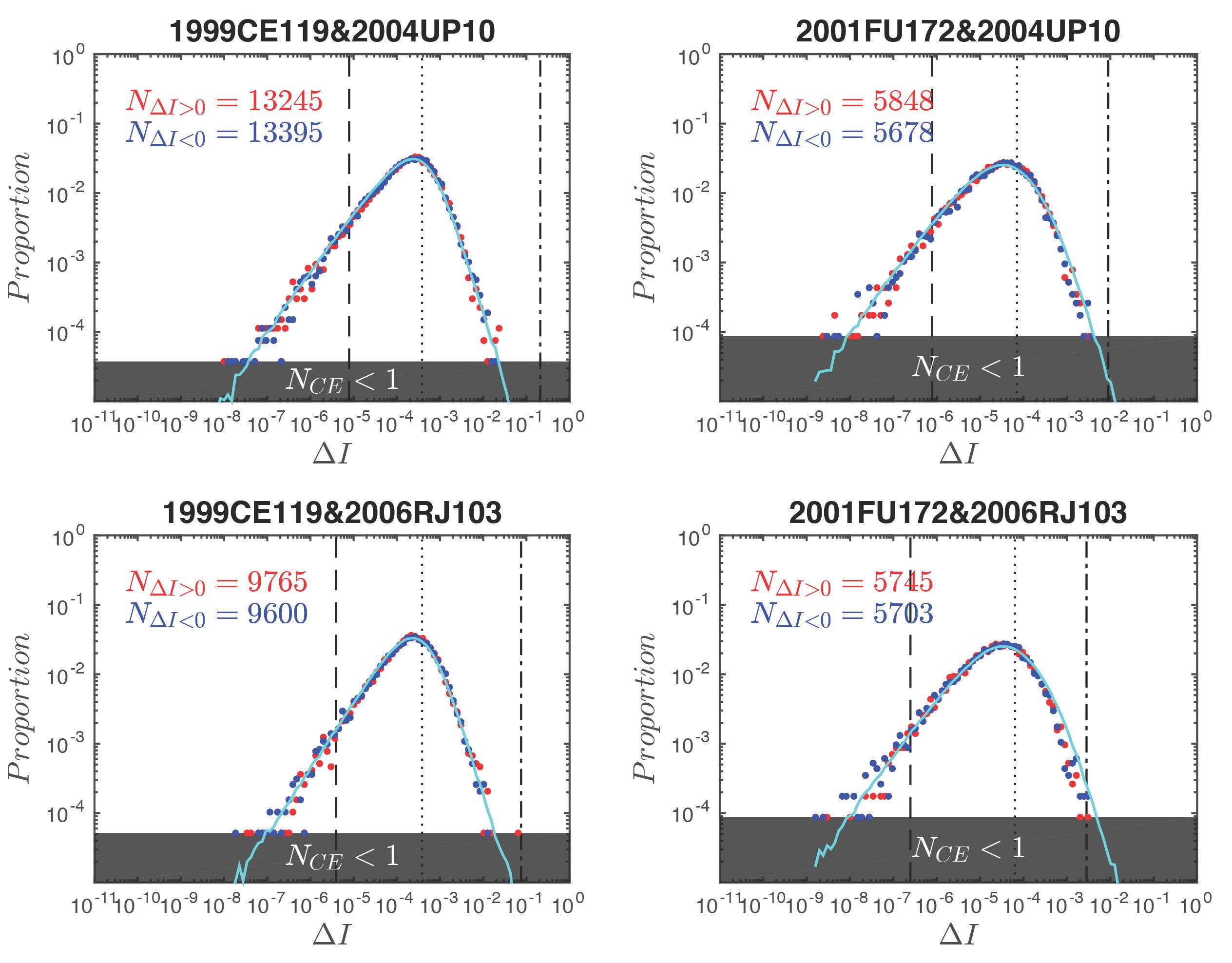}
      \caption{The distribution of $\Delta I$ in a logarithmic scale. The ordinate indicates the proportion to total number of CEs. The red and blue curves depict the positive and negative $\Delta I$ from numerical simulations respectively, with their corresponding number of CEs annotated on top. For each pair, the two curves overlap each other pretty well, implying a great symmetry between positive and negative effects. On each plot, there is a pronounced gap between the mean value and the mean absolute value, indicated by dashed and dotted lines respectively, which suggests a counteraction between positive and negative branches. The dash-dotted line indicates the absolute sum of all $\Delta I$. The numerical data are blocked by the shadow region, within which the proportion is too low to allow one single CE. The cyan curve depicts the theoretical result.}
    \label{Symba_diDstb_log}
    \end{figure}

Since we are trying to explore the interactions between Plutino and Trojan, the CE effect comes our main concern. Using the same data as in Sect. \ref{CElc}, we show the distribution of $\Delta I$, namely the inclination change of Trojan during each CE, in Fig.\,\ref{Symba_diDstb_log} in a logarithmic scale, with the positive and negative effect separated. The positive and negative CE effect conform to a nearly identical distribution, which implies that a CE has a completely even chance of increasing and decreasing the inclination of Trojan. Here we can introduce a two-sample K-S test to verify the consistency between positive and negative branches, with the result that the null hypothesis stands up for each pair of planetesimals. 

On the other hand, we notice some differences between the distributions from different pairs. Other than our intuition, a highly inclined Plutino in the right column in Fig.\,\ref{Symba_diDstb_log} leads to a relatively low CE effect, which is reflected by the lower mean absolute value denoted by the thick dashed line. This is closely linked to the terms associated with $\theta_0$ and $\alpha_v$ in the analytical expression of $\Delta I$ in Eq.\,(\ref{theoResult2}). Recall that in Fig.\,\ref{Symba_azmThetaDstb} the distribution of $\theta_0$ tends to concentrate on the center for Plutino with high inclination, which brings about relatively low values of the term $\sin\theta_0$. Meanwhile, the value of $\alpha_v$ will be conceivably closer to $0$ for Plutino with low inclination than high, since the Plutino and Trojan are more likely to orbit on a same plane, which leads to high value of the term $\cos\alpha_v$. On the whole, the two terms both tend to be lower for a more inclined Plutino.
    
    \subsubsection{The theoretical distribution of $\Delta I$}
    Actually with several derivations above combined together, a theoretical distribution of $\Delta I$ is available approximately. As mentioned before, Eq.\,(\ref{theoResult2}) only contains five variables, namely $\gamma_R$, $\theta_0$, $\alpha_v$, $\widetilde{\lambda_T}$ and $f_T$, where the distributions of the first two are analytically solved already. Besides, given the velocity vectors of Plutino and Trojan, the angle between the two, namely $\alpha_v$, can be easily derived, which gives
    \begin{equation}
    	\cos\alpha_v = \cos\theta_{v_P}\sin\varphi_{v_P}.
    \end{equation}

For the convenience of later analysis in the accompanying paper \citep{2017_2}, here we set a simplified approximation for $\alpha_v$. As shown in Fig.\,\ref{illu_thetavp}, the orbits of Plutino and Trojan can lie on the same side or different sides of the reference plane, leading to $\alpha_v$ either close to the sum or difference of their inclinations. Thus an average value roughly gives $\alpha_v \approx \left( (I_P+I_T)+|I_P-I_T|\right)/2 =\max(I_P,I_T)$.
   
     \begin{figure}[htbp]
    \centering
    \includegraphics[width=\hsize]{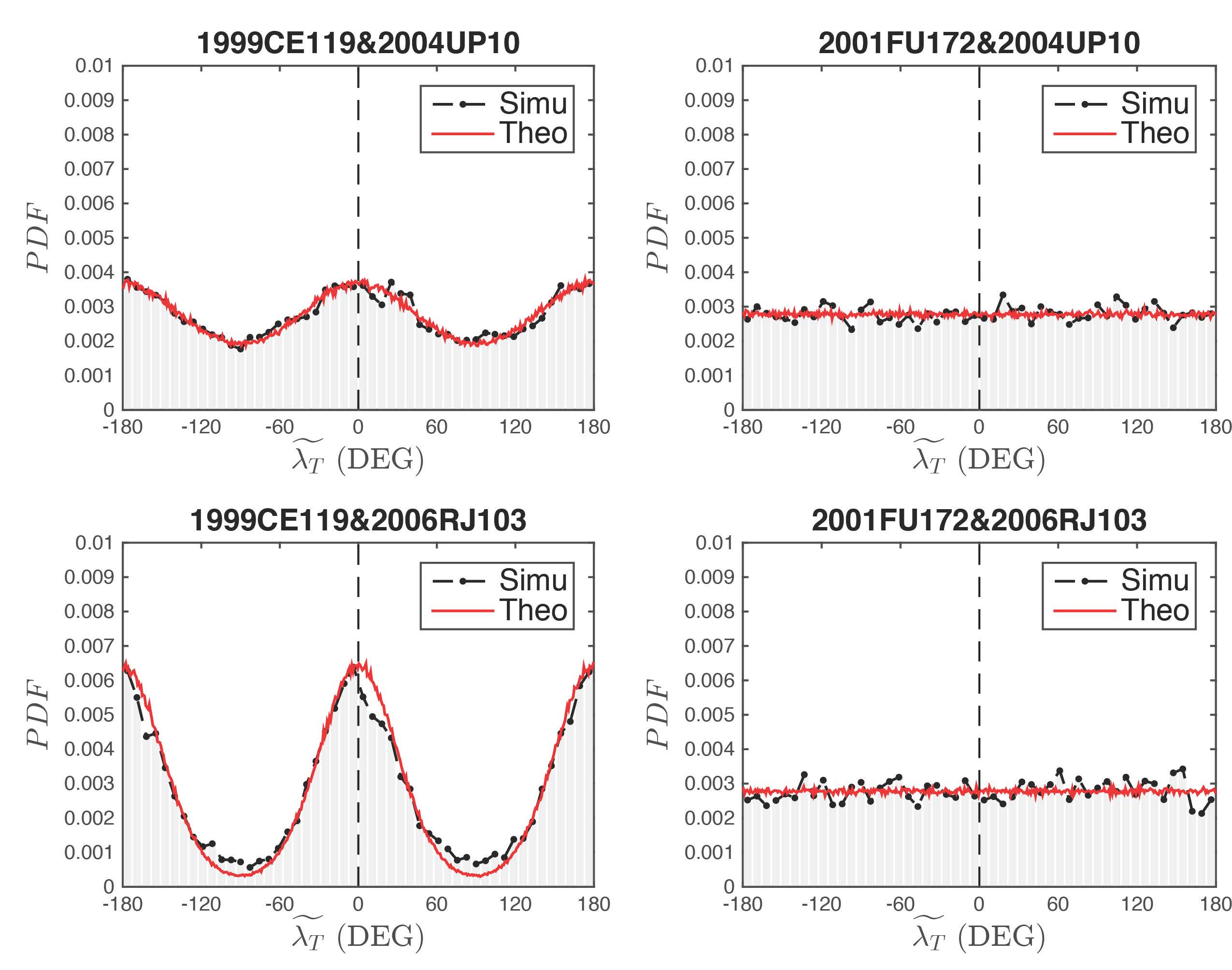}
      \caption{The probability density function of $\widetilde{\lambda_T}$. The dashed lines indicate $\widetilde{\lambda_T}=0\, \deg$, which coincide with the ascending node of Trojan. The highlighted lines are the theoretical predictions.}
         \label{Symba_azmLambdaDstb}
    \end{figure}
     
We now solve the distribution of $\widetilde{\lambda_T}$, which measures the argument of Trojan from its ascending node. This implies that $\widetilde{\lambda_T}$ relates to the specific position of Trojan, thus cannot be derived from the orientation of velocities in the co-moving system as the variables before. Nevertheless, the distribution of $\widetilde{\lambda_T}$ is still affected by the orbital characteristics of Putino and Trojan. 
     
Considering a model including a Trojan with high inclination and a Plutino orbiting on the ecliptic, i.e. with $0$ inclination, the Trojan is possible to meet Plutino only when it comes back to the ecliptic, namely at its ascending or descending node, in which case, the distribution of $\widetilde{\lambda_T}$ should concentrate on $0$ or $180\deg$. Another situation is that the Trojan has $0$ inclination while the Plutino has high inclination. In this case, the concentration occurs in the argument of Plutino, but the distribution of $\widetilde{\lambda_T}$ should be uniform due to the procession of Plutino. Other situations basically lie between the two. 
     
We now again apply the von Mises distribution to analytically estimate the distribution of $\widetilde{\lambda_T}$. Based on the above understanding, the distribution of $\widetilde{\lambda_T}$ should consists of two von Mises distributions locating on $0$ and $180\deg$ respectively, while the dispersion should be low when the inclination of Trojan is much higher than that of Plutino, and be high in reverse. Hence we can use $|I_T-I_P-I_C|$ to measure the dispersion, where $I_C$ is a preset bias. Using a linear correlation, we can empirically derive $\left(1/\sqrt{\kappa_f}\right)/\deg=8~|I_T/\deg-I_P/\deg-4|+25$.

The numerical and theoretical distributions of $\widetilde{\lambda_T}$ are together shown in Fig.\,\ref{Symba_azmLambdaDstb}, which are consistent with each other well.
     
    \begin{figure}[htbp]
    \centering
    \includegraphics[width=\hsize]{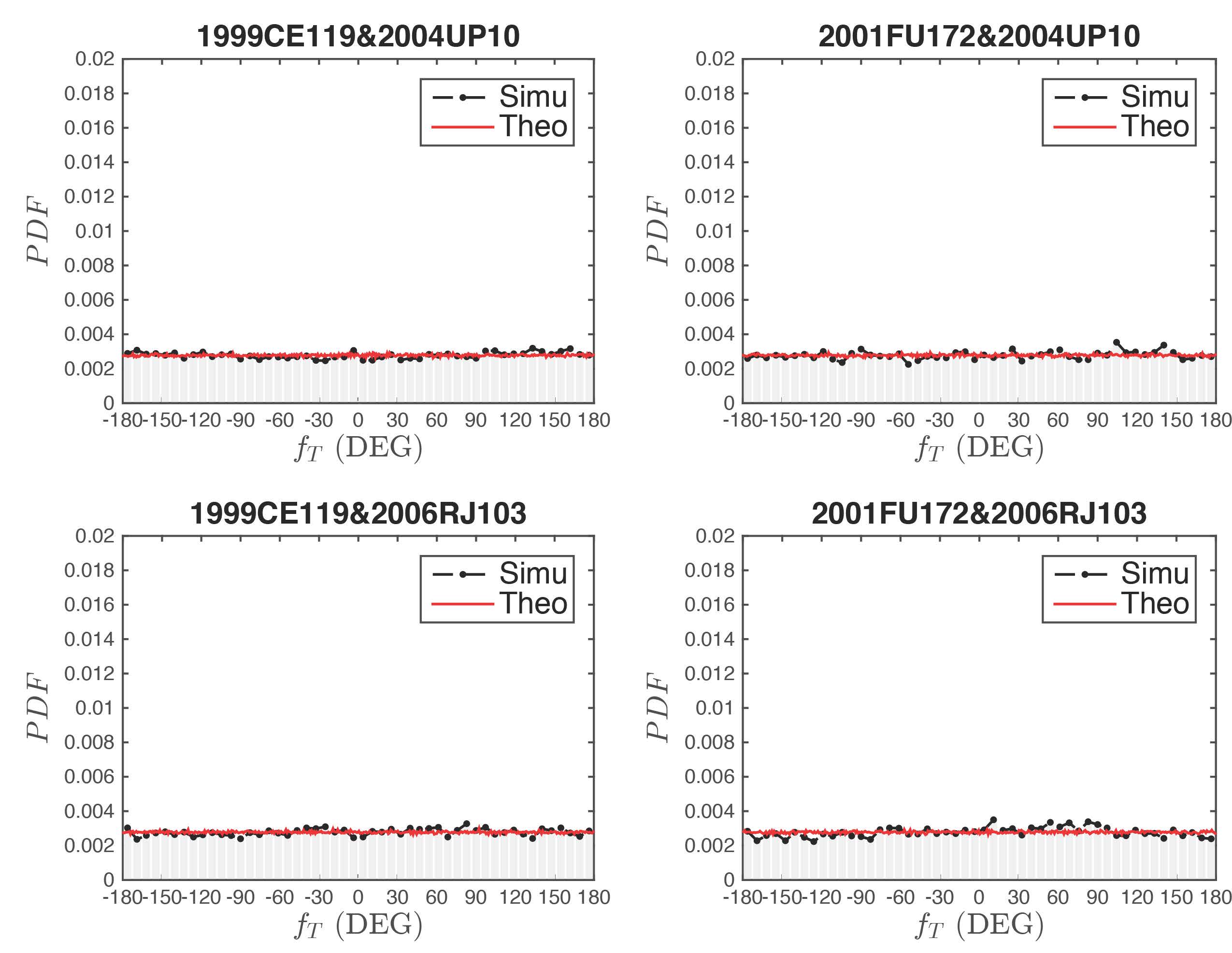}
      \caption{The probability density function of $f_T$. The highlighted lines are the theoretical predictions.}
         \label{Symba_azmfDstb}
    \end{figure}
    
The last involved variable is $f_T$, which we will simply treated as randomly distributed, since $f_T$ will not directly affect the location of CE given low eccentricity of Trojan. Fig.\,\ref{Symba_azmfDstb} supports this idea.
    
Finally we can obtain the theoretical distribution of $\Delta I$. In order to be consistent with the results from numerical simulations, we have to make a minor modification on the derivation of $\Delta I$. Recall that in the beginning of this section, we mentioned our method to calculate CE effect in the numerical simulation, which is simply the difference of the orbital element between the CE entry and exit. This method naturally forces the effect to be 0 if an encounter happens outside the CE region. Likewise, in the integration of Gaussian perturbation equation, we need to cut off the passage outside the CE region and only count the passage from the entry to the exit. This will not impact the statistics of the distribution of CE effect much since the contribution from the CEs outside the CE region is basically negligible. The modified expression of Eq.\,(\ref{theoResult2}) yields 
   \begin{equation}
   \label{theoResultRewrite}
   \Delta I = \Delta I_0\sqrt{\frac{1}{\gamma_R^2}-1},
   \end{equation}
where only the last term associated with $\gamma_R$ is different. All other variables remains unchanged and are incorporated into a dimensional coefficient for simplicity. We can clearly see that now as a CE approaches the boundary, i.e. $\gamma_R \to 1$, $\Delta I\to 0$ as expected.

Now given the distributions of other variables derived above, we obtain the theoretical distribution of $\Delta I$ through Eq.\,(\ref{theoResultRewrite}), which coincides with the numerical results in Fig.\,\ref{Symba_diDstb_log} fairly well.

In fact, if we focus on the key term $\gamma_R$, we can approximately derive an explicit expression for the distribution of $\Delta I$. Simply treat the absolute value of $\Delta I_0$ in Eq.\,(\ref{theoResultRewrite}) as constant and we have
   \begin{equation}
   \gamma_I  = \pm\sqrt{\frac{1}{\gamma_R^2}-1},
   \end{equation} 
where $\gamma_I\equiv \Delta I/|\Delta I_0|$ is a dimensionless variable denoting the magnitude of inclination change.

Based on this relation, Eq.\,(\ref{Dis2Dstb}) can be rewritten in the form with $\gamma_I$ serving as the independent variable, namely
   \begin{equation}
   P(\widetilde{\gamma_I}<\gamma_I) =
   \begin{cases}
   {\gamma_I}^2/\left(1+ {\gamma_I}^2\right), & \gamma_I \ge0,\\
   1/\left(1+{\gamma_I}^2\right), & \gamma_I<0 .
   \end{cases}
   \end{equation}
The probability density function of $\gamma_I$ can then be derived as 
   \begin{equation}
   \label{FSymba_diDstb_log}
   f(\gamma_I) = \frac{|\gamma_I|} {\left(1+{\gamma_I}^2\right)^{2}}.
   \end{equation} 
This functional form is quite close to Fig.\,\ref{Symba_diDstb_log}, with the probability density both leading to $0$ when $\gamma_I \to 0$ and $\gamma_I \to \infty$. The contributions from the angular terms are actually minor modifications to Eq.\,(\ref{FSymba_diDstb_log}).
    
    \begin{figure}[htbp]
    \centering
    \includegraphics[width=\hsize]{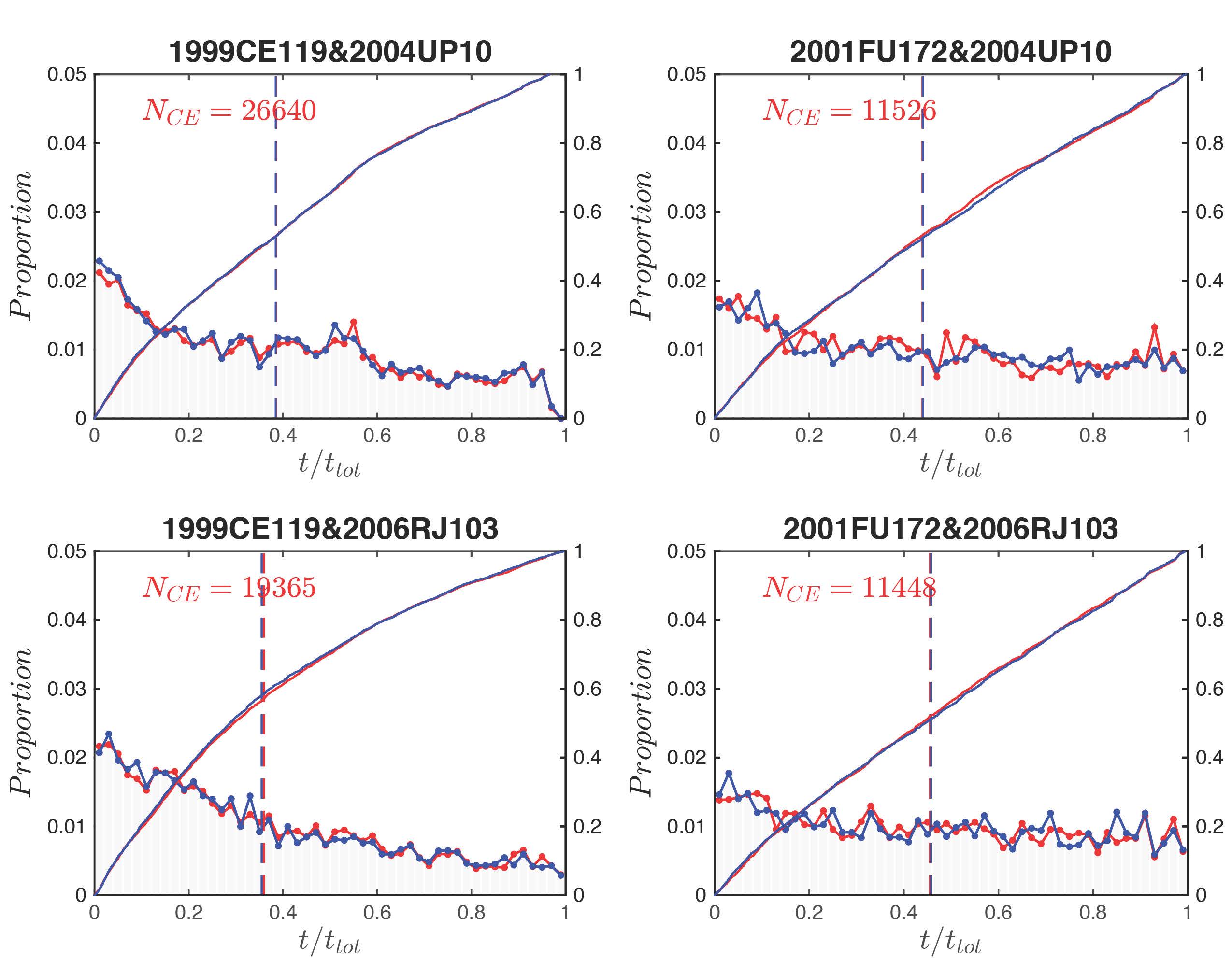}
      \caption{The respective frequency histograms of number of CEs with positive (Red) and negative (Blue) effects in same time intervals, as well as the corresponding ECDFs, similar to Fig.\,\ref{Symba_TimeDstb}. 
      The respective frequency histograms and ECDFs of positive and negative branches are both well consistent, revealing an equal rate of occurrence of two-sided effects. The vertical dashed lines indicates the respective mean value of positive and negative branch, which overlap each other well.}
    \label{Symba_TimeDstbPN}
    \end{figure} 
    
    \begin{figure}[htbp]
    \centering
    \includegraphics[width=\hsize]{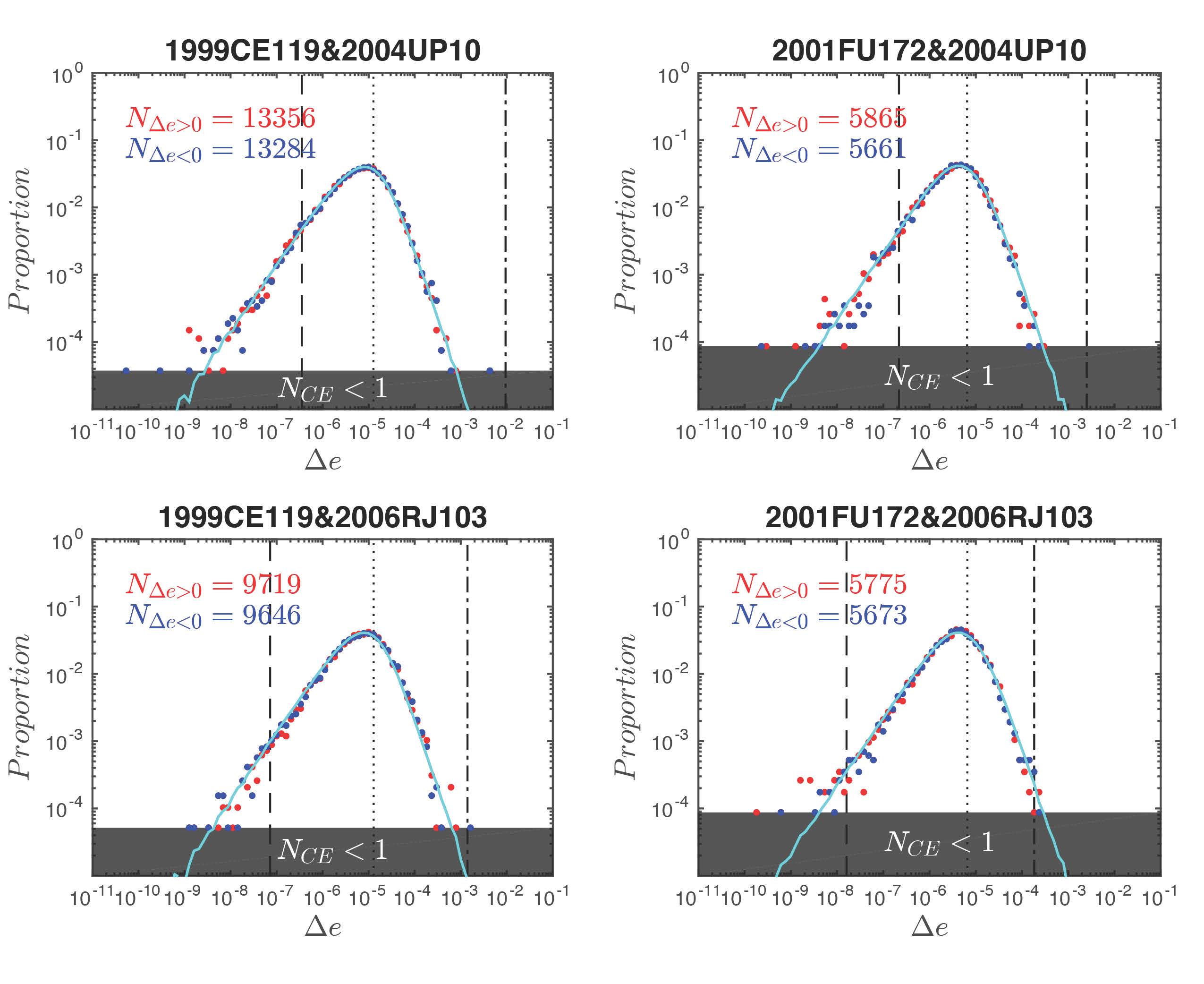}
      \caption{The distribution of $\Delta e$ in a logarithmic scale, similar to Fig.\,\ref{Symba_diDstb_log}. The cyan curve depicts the theoretical result.}
    \label{Symba_deDstb_log}
    \end{figure}
    
\subsubsection{The distribution of positive and negative CEs over time}
    Although the positive and negative branches are quite symmetrical in Fig.\,\ref{Symba_diDstb_log}, we worry about the concentration of biased CE effect within a particular time interval, which may cause a temporary influence despite the overall symmetry. Fig.\,\ref{Symba_TimeDstbPN} shows the number of positive and negative CEs over time in numerical simulations, as well as the empirical cumulative distribution function respectively. The curves drop with time because the clones we introduced may be unstable and gradually die out, as mentioned before. Nevertheless, our main interest is the comparison between the opposite effects, which barely shows any difference, pointing out a completely equal rate of occurrence of the two branches.
    
 \subsubsection{The distribution of $\Delta e$}
 The theoretical distribution of $\Delta e$ can be evaluated similarly, with no need for additional variables. As a reference, Fig.\,\ref{Symba_deDstb_log} shows the theoretical distribution of $\Delta e$ in cyan, which is consistent with the simulation data fairly well.
    
\section{Monte Carlo simulation}
\label{RanSim}
    We have paid great effort to demonstrate the randomness and impartiality lying in the close encounters between Plutinos and Trojans, which can be further verified by a Monte Carlo (M-C) simulation. The statistical distributions of previously discussed features will be reproduced here by the M-C method and juxtaposed with the numerical ones, to see if there is any difference between dynamic CEs and random ones.

\subsection{Method}
    \label{RanMethod}
Here we develop a M-C strategy to simulate fictitious CEs, which is briefly outlined as:

    \begin{enumerate}[Step 1]
    \item For each of the two planetesimals expected to CE, randomly choose a set of orbital elements within a given range, and convert into heliocentric coordinates. If the spatial distance between the two is less than a specific value, then a fictitious CE is obtained.
    \item Under the hypothesis that one body moves along a hyperbola against the other in a CE, the relative trajectory can be fitted based on the coordinates obtained above.
    \item Segment the hyperbolic trajectory within the CE sphere. In each segment use Gaussian perturbation equation to calculate the inclination or eccentricity change of Trojan, and add up to get the total effect of this CE.
    \end{enumerate}

\begin{figure}[htbp]
    \centering
    \includegraphics[width=\hsize]{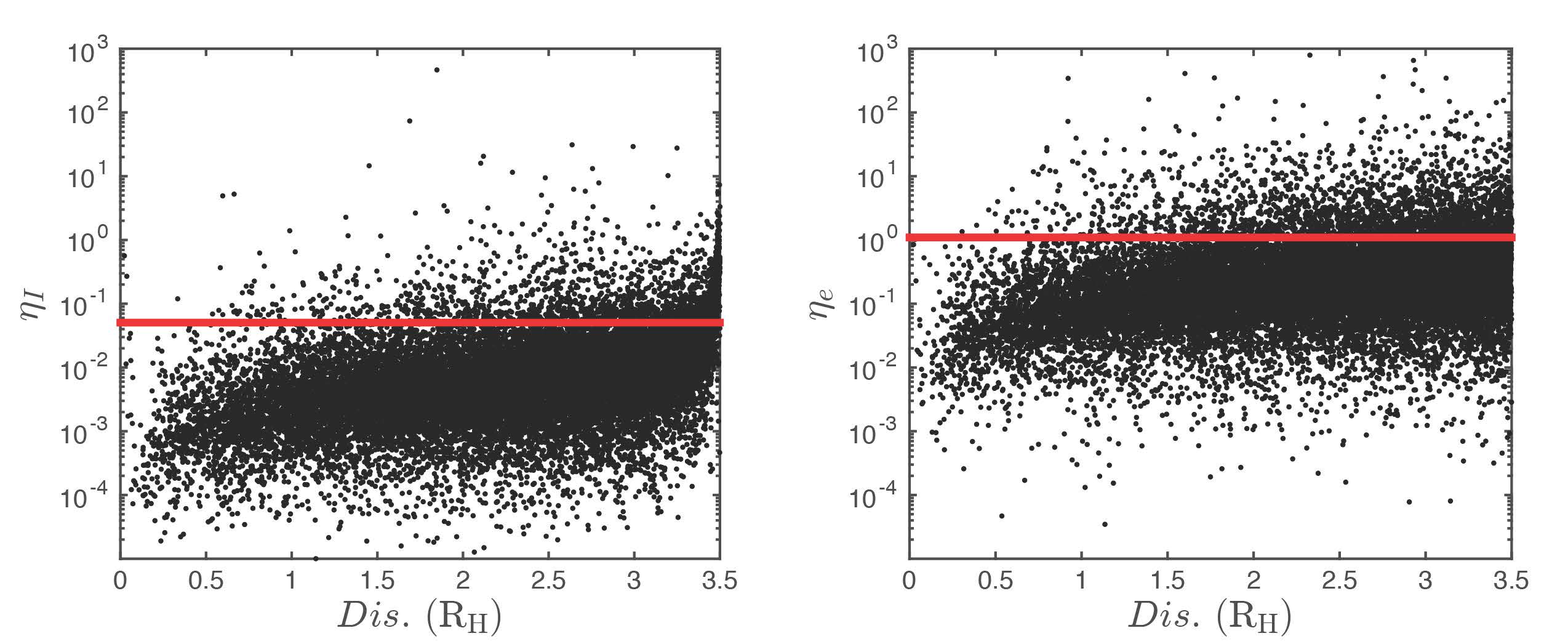}
      \caption{The relative error of the CE effect ($\Delta I$ and $\Delta e$) calculated in the Hyperbola-Perturbation way to that from numerical simulation. $\eta_I=\left |\left (\Delta I_{HP}-\Delta I_{NS}\right )/\Delta I_{NS}\right |$ and $\eta_e=\left |\left (\Delta e_{HP}-\Delta e_{NS}\right )/\Delta e_{NS}\right |$. Subscripts ``$HP$'' and ``$NS$'' correspond to the Hyperbola-Perturbation way and the numerical simulation respectively. The abscissa indicates the distance between the interacting particles. The highlighted line denotes the value below which $90\%$ of the dots reside.}
    \label{RanTestError}
    \end{figure}

Certainly the new strategy requires verification before being put in place. Apart from Step 1 that generates fictitious CEs, Step 2 - Step 3, summarized to be a Hyperbola-Perturbation method to calculate CE effect based on CE information, can be implemented to handle the CEs generated by a numerical simulation, with the results juxtaposed, to verify the practicability of this very method. Fig.\,\ref{RanTestError} shows the relative error of CE effect calculated in the Hyperbola-Perturbation way using the CE information already generated by a numerical simulation (1999 CE119\&2004 UP10 with $40$ Plutino clones), to the CE effect numerically obtained by this simulation. 
The result from the Hyperbola-Perturbation way are quite close to the experimental value except a relatively high deviation at a large distance, which is probably brought by the inaccuracy of hyperbolic fitting when the two particles are far away from each other. The disturbance of other planets may also acts since the perturbation way merely considers the influence from Plutino. Nevertheless, the peripheral CEs are after all less significant. Note that the relative error on eccentricity is remarkably higher than that on inclination, probably because the calculation of eccentricity change in the perturbation equation involves more variables, consequently bringing about additional errors. 
 
Note that in Step 3, given the fitted orbital elements of Trojan in each segment, we introduce the perturbation equation to derive the inclination and eccentricity change, rather than directly subtracting the inclination and eccentricity at the entry from that at the exit. In fact, the latter method, which seems more convenient, was also applied in our earliest attempts, but unfortunately deviating quite much from the numerical results. A possible explanation is that the prominent fitting error near the edge of the hyperbolic trajectory will be inherited or enlarged by the exit-entry method, while reduced by the perturbation method where the closest segments weigh the most.
    
\subsection{Results and comparison}
Now we can implement the M-C strategy to rapidly generate CE information. To present a direct comparison, here we will display simultaneously the results from numerical simulations and the M-C simulations on the statistical diagrams in Sect. \ref{Stat}. All the pairs of Plutinos and Trojans remain the same, as well as their initial conditions. The M-C simulations generate exactly the same number of CEs as produced by numerical simulation for each pair.
    
\subsubsection{Distribution of CE location}
\begin{figure}[htbp]
    \centering
    \includegraphics[width=\hsize]{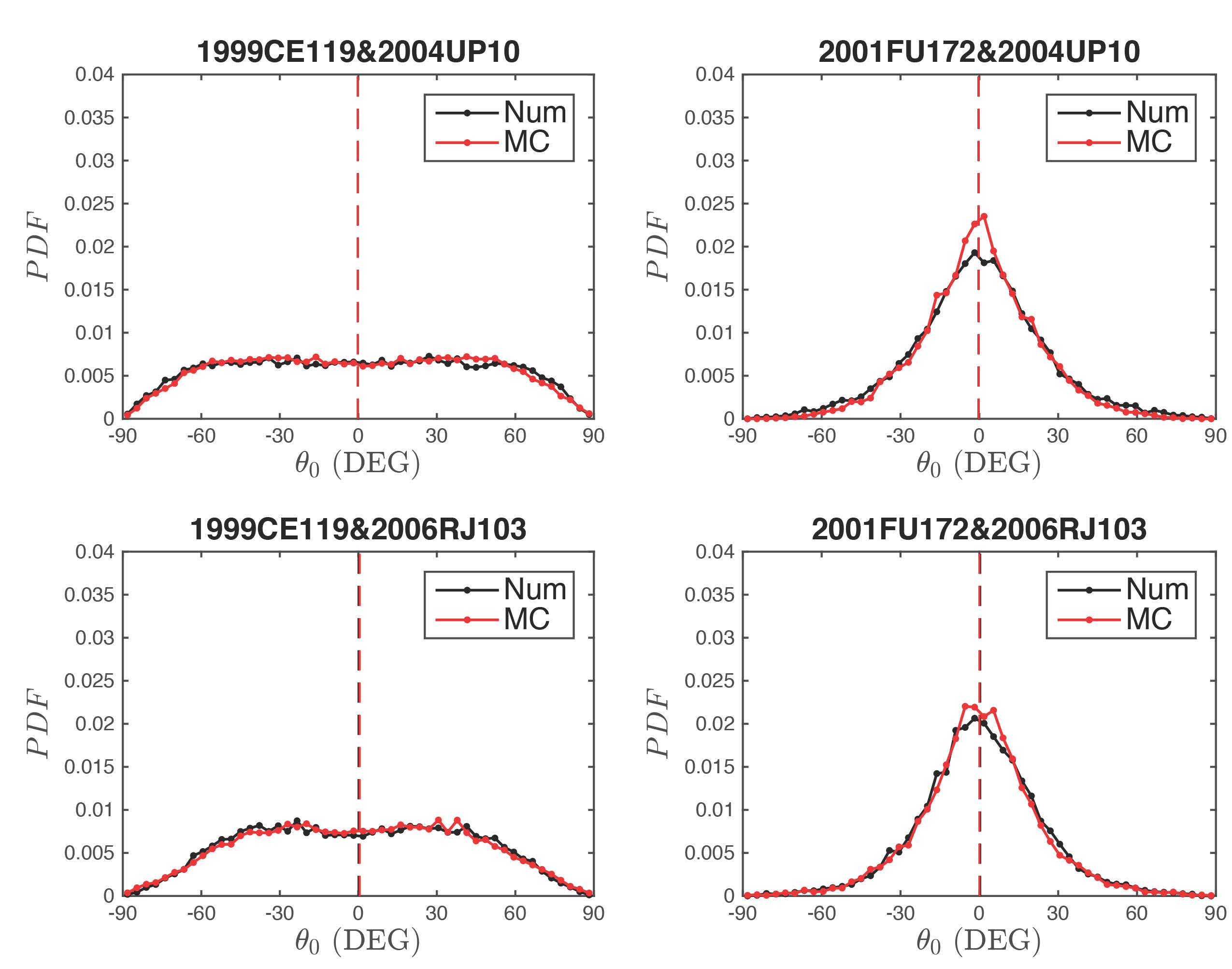}
      \caption{The probability density function of the latitude of the CE location $\theta_0$, 
      similar to Fig.\,\ref{Symba_azmThetaDstb}. The black is produced by the numerical simulation while the red is generated by the M-C simulation. The dashed line indicates the respective mean value.}
    \label{RanTestazmDstb}
    \end{figure}

\begin{figure}[htbp]
    \centering
    \includegraphics[width=\hsize]{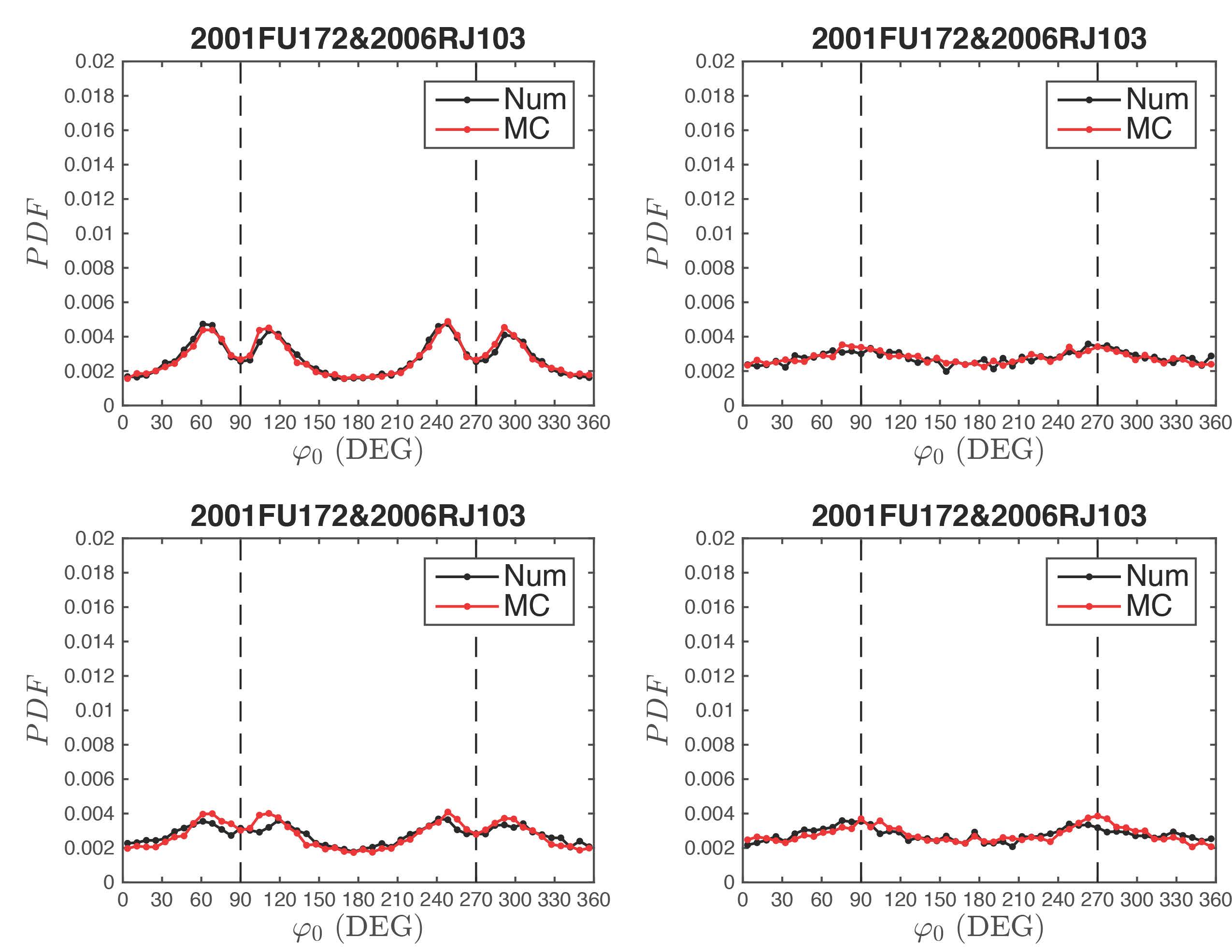}
      \caption{The probability density function of the longitude of CE location $\varphi_0$, similar to Fig.\,\ref{Symba_azmPhiDstb}. The black is produced by the numerical simulation while the red is generated by the M-C simulation.}
    \label{RanTestXazmDstb}
    \end{figure}

Similar to Figs.\,\ref{Symba_azmThetaDstb} and \ref{Symba_azmPhiDstb}, Figs.\,\ref{RanTestazmDstb} and \ref{RanTestXazmDstb} give a direct comparison between $\theta_0$, and $\varphi_0$ in CEs generated by numerical simulations and M-C simulations. In each figure, the two methods agree with each other fairly well. 
    
\begin{figure}[htbp]
    \centering
    \includegraphics[width=\hsize]{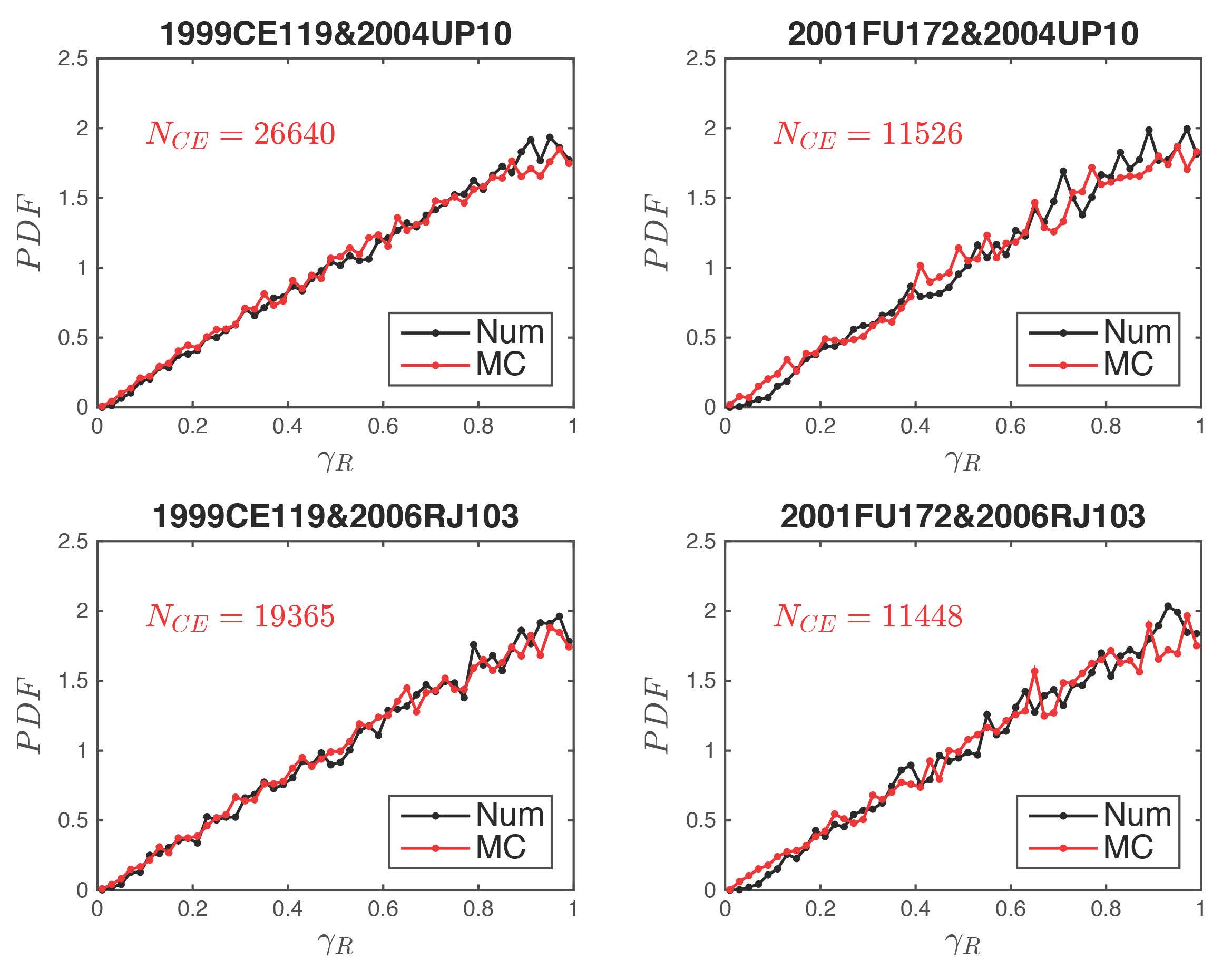}
      \caption{The probability density function of the minimum CE distance $\gamma_R$, similar to Fig.\,\ref{Stat_DisDstb4}. The black is produced by the numerical simulation while the red is generated by the M-C simulation.}
    \label{RanTestDisDstb}
    \end{figure}

Fig.\,\ref{RanTestDisDstb} gives the comparison of the minimum distance $R_0$ in CE between numerical simulations and M-C simulations, which are definitely consistent with each other, for the analytical linear distribution of $R_0$ is essentially derived from the uniform distribution of CE. 
    
To sum up, considering that $\theta_0$, $\varphi_0$ and $R_0$ are the key factors indicating the relative spatial position of the two planetesimals in CE, and by the fact that barely any difference has been found in the CE picture of a numerical simulation and a pure stochastic method, we can conclude that the occurrence of CE is generally a random event, with no particular bias, which again emphasizes our conclusion in Sect. \ref{Stat}. In turn, the consistency verifies that our M-C simulation is feasible in reflecting the details in CE. 
    
    \begin{figure}[htbp]
    \centering
    \includegraphics[width=\hsize]{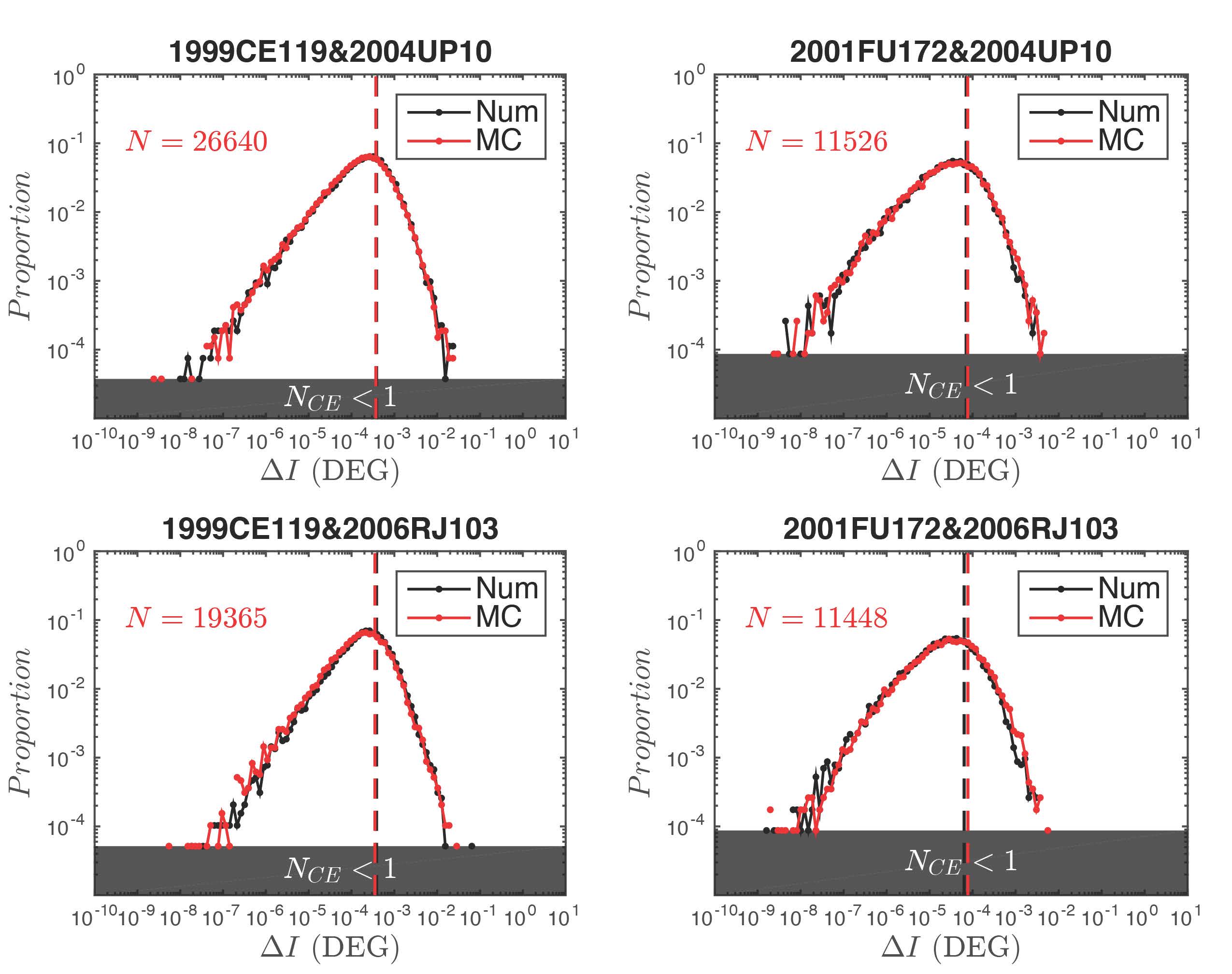}
      \caption{The distribution of absolute value of $\Delta I$ in a logarithmic scale. The ordinate indicates the proportion to total number of CEs. The black is produced by the numerical simulation while the red is generated by the M-C simulation. The dashed line indicates the absolute mean value.} 
    \label{RanTestdiDstb}
    \end{figure}
    
    \subsubsection{Distribution of CE time}
    
In a M-C simulation the concept of CE time is ambiguous since we are merely generating random numbers. However, the uniformity of the distribution of CEs over time in Sect. \ref{SectimeDstb} definitely agrees with the result of a stochastic method.
    
\subsubsection{Distribution of CE effects}
The vital criterion that determines the feasibility of the M-C simulation should be the distribution of CE effects, which is our major concern. Fig.\,\ref{RanTestdiDstb} compares the frequency histograms $\Delta I$ brought by numerical simulations and M-C simulations. The M-C productions basically coincide with the numerical data, and fully reproduce the main characteristics of the distribution curves, such as the correlation between high inclination and low CE effect. The mean value of the M-C productions is close to that of numerical data as well.

\section{Conclusion and discussion}\label{conclu}

In this work we implement a numerical method to efficiently detect close encounters (CEs) between Plutinos and Neptune Trojans. For typical planetesimals in reality, detailed statistical analyses are performed on the frequency and geometries of CEs, followed by consistent analytical estimations. We hereby present a better understanding of the CE picture between planetesimals in overlapping resonances, and reveal the symmetry and stochasticity of such orbital communications. 

Specifically, as covered in Sect. \ref{CElc}, the CEs are found to distribute symmetrically against the orbital plane of Trojan inside the CE region, thus in whole exerting unbiased influence on Trojan. The minimum CE distance well conforms to a linear distribution, consistent with a ``cross section" picture. Over a long timespan, the CEs take place uniformly, with no tendency of concentration. Particular orbital characteristics, typically the inclination, will directly affect the frequency of CE, as well as the angular distributions of CEs such as the longitude, latitude inside the CE region and the argument from the ascending node on the orbit of Trojan. Such features can be well explained by the approximate theory.

Besides, we investigate the CE effect, measured by the inclination and eccentricity change of Trojan. By detailedly comparing the distribution of positive and negative effects, along with their respective frequency over time, we conclude that the CE effect is impartial, thus little possible to observably alter the orbital elements of Trojan. In other words, the high inclination problem of Trojan is unrelated to the communications contributed by Plutinos. Furthermore, the theoretical distribution of CE effect is derived from the the analytical estimations on associated variables according to Eq.\,(\ref{theoResult2}) and Eq.\,(\ref{theoResulte}), namely the integral formula of Gaussian perturbation function. Since the terms like $\cos{\widetilde{\lambda_T}}$ and $\sin\theta_0$, which directly determine the sign of CE effect, are all unbiased due to the symmetrical distribution of $\widetilde{\lambda_T}$ and $\theta_0$, the theoretical effect is certainly impartial, which again verify the numerical result. 

Though Plutinos contribute little to the characteristics of Trojans by CEs, they may alternatively take effect by more violent interactions, namely collisions. Actually we can estimate the collision rate between Plutinos and Trojans for a regular case, using Eq.\,(\ref{Dis2Dstb}), namely the linear distribution of $R_0$. 
Taking Pluto as an example, $R_{th}\approx0.18\,\rm{AU}$, while naturally the minimum distance $R\lesssim 2R_{Pluto}$ for a collision caused by a sizable Trojan, where the radius of Pluto $R_{Pluto}\approx1000\,\rm{km}$.  As a result, the normalized collision probability is estimated as $P\approx2\times 10^{-9}$. Besides, in numerical simulations the number of CEs between one Plutino and one Trojan is generally the order of $10^3$, which implies that averagely only one collision will happen given $5\times10^5$ Trojans during $1$Gyr. Despite the large quantity of potential Neptune Trojans in reality, this is an extremely low probability, not to mention that typical planetesimals are far smaller than Pluto. Consequently, the collisions brought by Plutinos are too scarce to cause prominent effects on the overall distribution features of Trojans.
 
Inspired by the above results, we can infer that Trojans may in turn have finite influence on the distribution features of Plutinos, no matter by CEs or by collisions. The latter comes obvious because the collision rate is a mutual factor for Plutino and Trojan. The former is tenable because Trojan should exert impartial effects on Plutino as well since the opposite case is true. The numerical result confirms this inference, but is further found to introduce slightly higher CE effect than the opposite case.
This may be theoretically explained by the variations of the integral formulas, i.e. Eq.\,(\ref{theoResult2}) and Eq.\,(\ref{theoResulte}), where the pertinent variables are substituted with that related to Plutino, when the effect on Plutino is considered.

Till now plenty of attention is drawn on the change of orbital elements including inclination and eccentricity, but one may notice that, there is no mention of the potential CE effect on the semi-major axis. The reason lies that the issue will go beyond the scope of our model associated with resonant planetesimals, if the semi-major axis, which strictly restricts the boundary of resonance, undergoes prominent change. In fact, the semi-major axis changes of Trojan in our simulations turn out to be quite small, otherwise Trojan will have no chance to keep staying in resonance during almost every simulation. 

In the final part of this work, we implement a Monte Carlo (M-C) simulation to stochastically generate CEs, along with a Hyperbola-Perturbation method to calculate consequent effects. The consistency of statistical diagrams between numerical integrations and M-C simulations again proves the randomness and impartiality of CEs between Plutinos and Trojans.
In addition, since the CE productions by M-C simulations are close to that by numerical integrations, we can use that to generate CEs in place of numerical approach under specific circumstances, with the benefit of high computational efficiency and no restriction of number of CEs. 
      
We will further discuss the cumulative effect of CEs on the orbit of Trojan, for fear that tiny random effect may accumulate to be prominent, as long as there is a tremendous number of CEs or a sufficient number of Plutinos simultaneously interacting. That is where the theoretical distributions of CE effects come in handy. Though derived from the pattern of 2+2 planetesimals, we shall show in the accompanying paper that this analytical tool is actually applicable to a wide range of orbital elements. Therefore, with the help of the random walk theory, we can easily estimate the cumulative effect of CEs contributed by an arbitrary Plutino, or even a group of different Plutinos. In this way, we are able to give the possible range of CE effects brought by realistic Plutinos, much more convincing than the result of pure numerical simulations including even hundreds of planetesimals.
      
In the end, it is necessary to point out that, our statistical methods and analytical estimations developed in this work can be directly applied to other circumstances associated with the overlap between mean motion resonances, e.g. Hilda group and Jupiter Trojans. Once the randomness of CEs is identified, further theoretical tools introduced in the subsequent paper can be implemented to quantitively determine the potential effect. Specific results may differ from what obtained for the case introduced in this work, due to distinct orbital and physical characteristics. Of course the CEs between planets and comets can be studied in the same way, but generally the low frequency and large effect in one single CE may impair the statistical significance.

\acknowledgments

The authors wish to thank Prof. Dai Xiongping for an inspiring talk on related subjects. The appreciation also goes to the editor and the anonymous referee for their constructive comments that helped to improve the paper. This work has been supported by the National Natural Science Foundation of China (NSFC, Grants No.11473016 \& No.11333002).

\bibliographystyle{aasjournal} 
\bibliography{APJ.bib} 
%
  
\end{document}